\documentclass[
    aps,            
    prd,            
    twocolumn,      
    superscriptaddress, 
    nofootinbib,    
    amsmath,        
    amssymb, 
   longbibliography 
]{revtex4-2}

\usepackage{graphicx}  
\usepackage{dcolumn}   
\usepackage{bm}        
\usepackage{color}     
\usepackage{subcaption}
\usepackage[
    colorlinks=true,
    linkcolor=blue,
    citecolor=blue,
    urlcolor=blue,
    bookmarks=true,
    bookmarksnumbered=true
]{hyperref}
\usepackage[margin=0pt, font=small, labelfont=bf, justification=justified]{caption}

\begin{document}

\title{Static Dark Fluid Thin Shells in Schwarzschild-de Sitter Spacetimes: Stability and Black Hole Shadows}

\author{Dimitrios Efstratiou}
\email{d.efstratiou@uoi.gr}
\affiliation{Department of Physics, University of Ioannina, GR-45110, Ioannina, Greece}

\author{Evangelos Achilleas Paraskevas}
\email{e.paraskevas@uoi.gr}
\affiliation{Department of Physics, University of Ioannina, GR-45110, Ioannina, Greece}

\author{Leandros Perivolaropoulos}
\email{leandros@uoi.gr}
\affiliation{Department of Physics, University of Ioannina, GR-45110, Ioannina, Greece}

\date{\today}
\begin{abstract}
We study the existence and radial stability of static, spherically symmetric thin shells joining two Schwarzschild--de~Sitter (SdS) spacetimes $(m_\pm,\Lambda_\pm)$. Using the Israel junction formalism, we map the stable equilibria ($V_{\mathrm{eff}}''>0$) of the effective potential. Near the equilibrium radius $R_0$ the shell's surface density $\sigma$ and pressure $p$ obey the linearized barotropic law $p=p_0+c_s^2(\sigma-\sigma_0)$, with sound speed $c_s^2=\lambda c^2$. Since $c_s^2$ is independent of the equilibrium ratio $w_0\equiv p_0/(\sigma_0 c^2)$, tension shells ($w_0<0$) stay radially stable with real $c_s$. Fixing  $\Lambda_+$ so that its vacuum energy density equals the critical density (Planck~2018), and taking $m_-$ representative of astrophysical black holes, we systematically map the stable equilibria $(R_0,\sigma_0)$ over $(m_\pm,\Lambda_\pm,\lambda,w_0)$ and find that stable shells with $\sigma_0>0$ and $0<\lambda\leq 1$ exist only for $m_+/m_->1$, at three scales---the photon sphere, the SdS static radius, and the cosmological horizon. At $\lambda=1$ the numerical windows, checked against the analytic test-shell bounds, are $(1-\sqrt{13})/6\lesssim w_0\lesssim 1/2$ ($\Lambda_+=\Lambda_-$), $-2/3\lesssim w_0\lesssim 1/2$ ($\Lambda_+>\Lambda_-$), and $0\lesssim w_0$ ($\Lambda_+<\Lambda_-$). Positive-pressure shells ($0\lesssim w_0\lesssim 1/2$) may sit near the photon sphere for any ordering of $\Lambda_+/\Lambda_-$. Additionally, when $\Lambda_+<\Lambda_-$ with $w_0\gtrsim0$, they may also sit at the static-radius scale. Tension shells are absent for $\Lambda_+<\Lambda_-$; for $\Lambda_+=\Lambda_-$, they reach the cosmological horizon scale; for $\Lambda_+>\Lambda_-$, they are confined to the static-radius scale. Finally, we compute the dark fluid shell's imprint on the SdS black-hole shadow seen by a static observer at varying radial distance; the radially stable tension shells of interest sit near the static radius, where the deviation likely lies beyond observational reach.
\end{abstract}

\maketitle

\section{Introduction}
\label{sec:intro}
 
Thin shells in static, spherically symmetric spacetimes idealize gravitational collapse~\cite{LeMaitre:2019xez}, domain walls~\cite{PhysRevD.23.852}, and braneworlds~\cite{PhysRevLett.83.3370,Antoniadis:1998ig,Arkani-Hamed:1998jmv}. Israel's junction formalism~\cite{Israel:1966rt} glues two manifolds across a timelike hypersurface with continuous induced metric, the extrinsic-curvature jump fixing the surface stress-energy~\cite{Kijowski2006,Poisson:2009pwt,GronHervik2007GR}; an exact static thin-shell solution around a black hole~\cite{Frauendiener:1990nao} was later analysed for stability~\cite{Brady:1991np,Schmidt:1999wc}. Such models also describe stellar structures and gravastars~\cite{Zloshchastiev:1998gv,Chan:2009hp,Pereira:2014lta,Pradhan:2023wac,Visser:2003ge,Pani:2009ss} and cosmological bubbles~\cite{Blau:1986cw,BEREZIN198391,PhysRevD.34.2913,Berezin:1987bc}, their stability~\cite{Poisson:1995sv,Ishak:2001az,Goncalves:2002yf,Visser:2003ge,Lobo:2005zu,HabibMazharimousavi:2017zlc,Alestas:2020wwa,Masa:2020hok,Cataldo:2025ukr} and collapse~\cite{Crisostomo:2003qc,Crisostomo:2003xz,Bueno:2024zsx,e28010096} studied extensively, with extensions to alternative gravity~\cite{Gravanis:2007ei,Javed:2024trj}, scalar-field~\cite{Beauchesne:2011au,Javed:2023rdf,Hussain:2024vqc}, rotating~\cite{Musgrave:1995ka,Gleiser:2008du,Antoniou:2022vxi}, and charged~\cite{Eiroa:2011nd,Reyes:2022xaj} configurations, charge screening~\cite{HabibMazharimousavi:2013qvv,Mazharimousavi:2015sga}, lower-dimensional thermodynamics~\cite{Lemos:2013dsa,Lemos:2014eva}, and cloud-of-strings geometries~\cite{Eiroa:2016rdd,HabibMazharimousavi:2017zlc}. The cut-and-paste procedure yields thin-shell wormholes supported by exotic matter at a throat~\cite{Poisson:1995sv,Ishak:2001az,Lobo:2003xd,Lobo:2004uq,Lobo:2004id,Eiroa:2007qz,Lemos:2008aj,Dias:2010uh,Garcia:2011aa,Nakao:2013hba,Sharif:2013xta,Tsukamoto:2018lsg,Li:2018jxy,Amirabi:2019uva,Berry:2020tky,Lobo:2020vqh,Lobo:2020kxn,Sharif:2021cvr,Rosa:2023olc,Eid:2024iza,Eid:2023fbh,Lobo:2025tph,Rippentrop:2025dfa,Antoniou:2022dre}. More broadly, the imprints of thin-shell configurations on strong-field lensing~\cite{Perivolaropoulos:2023zzc} and on black-hole shadows and photon rings~\cite{Sakai:2014pga,Antoniou:2022dre,Peng:2021osd,He:2024yeg,Cao:2025qzy,Saleem:2025axo,Macedo:2025ipc,Laeuger:2025zgb,Wu:2024juj,BenAchour:2025uzp,Konoplya:2025mvj} have recently drawn attention.

We study the existence and radial stability of static, spherically symmetric thin shells separating two Schwarzschild--de~Sitter (SdS) regions with possibly distinct masses $m_\pm$ and cosmological constants $\Lambda_\pm$. We consider genuine two-sided surface layers with outward-pointing normals, not cut-and-paste wormholes (whose inward normals require exotic matter)~\cite{KIM199213,Ishak:2001az,Lobo:2003xd,Alestas:2020wwa,Eid:2024iza,Rippentrop:2025dfa,e28010096}. The shell is characterized by a surface density \(\sigma\) and an isotropic pressure \(p\), which obey the equation of state \(p = \lambda (\sigma - \sigma_1) c^{2}, \quad \sigma_1 = (\lambda - w_0)\sigma_0/\lambda\), or equivalently, a first-order Taylor expansion of a generic barotropic relation \(p(\sigma)\) around the equilibrium point \((\sigma_0, p_0)\): \(p \simeq p_0 + \left.\frac{\mathrm{d}p}{\mathrm{d}\sigma}\right|_{\sigma_0} (\sigma - \sigma_0) = p_0 + c_s^{2} (\sigma - \sigma_0)\). Here \(c_s^{2} = \mathrm{d}p/\mathrm{d}\sigma = \lambda c^{2}\) (with $0\le\lambda\le1$ ensuring a real sound speed), while the equilibrium values satisfy \(p_0 = w_0 \sigma_0 c^{2}\), with \(\sigma_0 \equiv \sigma(R_0)\), \(p_0 \equiv p(R_0)\), and \(w_0 \equiv p_0/(\sigma_0 c^{2})\). The single-parameter law $p=w_0\sigma c^{2}$ is the special case in which the expansion runs through the origin ($\sigma_1=0$, $\lambda=w_0$), tying $c_s^{2}=w_0c^{2}$ and giving imaginary $c_s$ for tension shells ($w_0<0$)~\cite{Brady:1991np,Babichev:2004qp,Babichev:2004yx,Paraskevas:2025shf}. Because $\lambda$ is independent of $w_0$, even tension shells ($w_0<0$) keep a real sound speed, and these stable negative-pressure layers are our main focus.

We treat the shell as a \emph{dark structure}: it does not couple to the electromagnetic field, so it is transparent and interacts only gravitationally, and we idealize it at a single scale $R_0$. We make no claim it represents a specific object; rather, these are toy models that may capture some qualitative features of familiar configurations. The mass step ($m_+>m_-$, $\Lambda_+=\Lambda_-$) idealizes, as a single layer, a dark-matter overdensity around a black hole~\cite{Bertone:2024rxe,Peebles1972,Young1980,Dehnen:1993uh,Quinlan:1994ed,Gondolo:1999ef,Sadeghian:2013laa}, while a jump $\Lambda_+\neq\Lambda_-$ models a domain or bubble wall between regions of different vacuum energy~\cite{Zeldovich:1974uw,Kibble:1976sj,Coleman:1977py,Callan:1977pt,Coleman:1980aw,Vilenkin:1984ib,Dong:2011gx,Babichev:2021uvl,Wei:2022poh}. A closely related geometry appears in the Brown--Teitelboim mechanism \cite{Teitelboim:1985ya,Brown:1987dn,Brown:1988kg}, where a closed membrane charged under a three-form potential separates two vacuum regions whose cosmological constants differ by one unit of the membrane charge \cite{Bousso:2000xa,Feng:2000if}; the branch lowering $\Lambda$ inside the membrane is the more probable one \cite{Brown:1988kg}. A central mass is no obstruction, membranes between two Schwarzschild--de~Sitter regions of distinct masses having been studied in Ref.~\cite{PhysRevD.69.083520} with the same junction condition. The three orderings ($\Lambda_+>\Lambda_-$, $\Lambda_+=\Lambda_-$, $\Lambda_+<\Lambda_-$) are treated on the same footing, with $\Lambda_+=\Lambda_-=0$ recovering matched Schwarzschild spacetimes.

This setup poses two questions: \emph{where can such a layer sit stably, and how large is its shadow deviation?} The forward problem is clean: fixing $(m_\pm,\Lambda_\pm,\lambda,w_0)$ selects the stable equilibrium $(R_0,\sigma_0)$ and hence a definite deviation. Our contribution is threefold. (i) We treat two-sided SdS layers with independently distinct $m_\pm$ and $\Lambda_\pm$, closed by the equation of state, which admits stable tension shells with real $c_s$. (ii) Fixing the exterior cosmological constant  $\rho_{\Lambda_+}=\rho_{\mathrm{crit}}=3H_0^2/(8\pi G)$, with $H_0$ at its \emph{Planck}~2018 value~\cite{Planck2018} and varying $\Lambda_-$ at the same scale, we map the stable equilibria $(R_0,\sigma_0)$ over $(m_\pm,\Lambda_\pm,\lambda,w_0)$, showing they exist only for $m_+/m_->1$ and at three characteristic scales---the photon sphere, the SdS static radius $R_{\mathrm{st}}=(3Gm_-/c^2\bar\Lambda)^{1/3}$(a characteristic
scale for the matched configuration, introduced for convenience as the
reference length at this scale and carrying no further significance; it is not the static radius of either
patch, see Appendix~\ref{sec:appendixa}), and near the cosmological horizon; the three $w_0$ windows are fixed by whether $\Lambda_+/\Lambda_-$ exceeds, equals, or falls below unity. (iii) We compute the imprint of such a transparent layer on the SdS black-hole shadow~\cite{Perlick:2021aok,PhysRevD.60.044006,Perlick:2018iye} for a static observer. Section~\ref{stability} sets up the junction conditions, effective potential, and stability windows (analytic and numerical); Section~\ref{sec:signatures} studies the shadow signatures; Section~\ref{sec:conclusion} summarizes the implications. Detailed derivations are collected in Appendices~\ref{app:junction}--\ref{sec:appendixa}.

\section{Radial Stability of Static Thin Shells in Schwarzschild--de Sitter Geometry}
\label{stability}
 
Examples of static, stable thin shells or wormholes in SdS or anti--de~Sitter backgrounds appear in Refs.~\cite{KIM199213,Ishak:2001az,Lobo:2003xd,Eid:2024iza,Rippentrop:2025dfa,e28010096}. Let $\Sigma\subset\mathcal{M}$ be a timelike hypersurface separating two SdS regions, $\mathcal{M}=\mathcal{M}^{+}\cup\mathcal{M}^{-}$ with $\partial\mathcal{M}^{+}\cap\partial\mathcal{M}^{-}=\Sigma$, the two sides carrying distinct $m_\pm$ and $\Lambda_\pm$. Each side is static, spherically symmetric, and satisfies $R^\pm_{\mu\nu}=\Lambda_\pm g^\pm_{\mu\nu}$, with\footnote{We retain explicit factors of $G$ and $c$ for clarity.}
\begin{equation}\label{metricSdSmain}
ds_{\pm}^2= c^2 f_\pm(r_\pm)\,dt_\pm^2- \frac{dr_\pm^2}{f_\pm(r_\pm)}
- r_\pm^2\bigl(d\theta^2+\sin^2\theta\,d\phi^2\bigr),
\end{equation}
\begin{equation}\label{frmetricmain}
f_\pm(r_\pm)=1-\frac{2Gm_\pm}{c^2 r_\pm}-\frac{\Lambda_\pm}{3}r_\pm^2 .
\end{equation}
Adopting intrinsic coordinates $y^{i}=(c\tau,\theta,\phi)$ with $\tau$ the comoving proper time, continuity of the induced metric across $\Sigma$ forces $R_+=R_-\equiv R$. The Israel junction conditions relate the jump in extrinsic curvature to the surface stress--energy tensor $S_{ij}$~\cite{GronHervik2007GR},
\begin{equation}\label{israeljunc0}
[\mathcal{K}_{ij}]_\pm=\frac{8\pi G}{c^4}\left(S_{ij}-\tfrac12 h_{ij}S\right),
\qquad [X]_\pm\equiv X_+-X_- ,
\end{equation}
with $S=h^{ij}S_{ij}$. The form of $S_{ij}$ is not assumed but follows from spherical symmetry: the induced extrinsic curvature is diagonal with $\mathcal{K}^\pm_{\phi\phi}=\sin^2\!\theta\,\mathcal{K}^\pm_{\theta\theta}$, so $S_{ij}$ is necessarily of perfect-fluid form,
\begin{equation}\label{perfectfluid}
S_{ij}=\bigl(\sigma+p/c^2\bigr)v_i v_j - p\,h_{ij},
\end{equation}
where the surface density $\sigma$ and isotropic pressure $p$ are fixed by the $\tau\tau$ and angular components of \eqref{israeljunc0}. The  embedding, four-velocity, outward normal, and extrinsic-curvature computation are given in Appendix~\ref{app:junction}. The $\theta\theta$ component yields the junction condition on $\Sigma$,
\begin{equation}\label{junc1main}
\sqrt{f_-+(\dot R/c)^2}-\sqrt{f_++(\dot R/c)^2}=\frac{4\pi G}{c^2}\,\sigma(R)\,R ,
\end{equation}
 while conservation of the surface stress--energy, $\hat\nabla_i S^{ij}=0$, has a single nontrivial ($j=\tau$) component~\cite{Visser:2003ge}
\begin{equation}\label{conservationofenergysigma1}
\dot\sigma+\frac{2\dot R}{R}\bigl(\sigma+p/c^2\bigr)=0 .
\end{equation}
Writing \eqref{junc1main} as~\cite{Brady:1991np}
\begin{equation}\label{betaeqm}
\beta_- - \beta_+ = \kappa(R),
\end{equation}
\begin{equation}\label{kapparmain}
\beta_\pm\equiv\sqrt{f_\pm+(\dot R/c)^2},\qquad
\kappa(R)\equiv\frac{4\pi G}{c^2}\,\sigma(R)\,R ,
\end{equation}
factorising $\beta_-^2-\beta_+^2=f_-(R)-f_+(R)$ gives $\beta_\pm=\tfrac{f_--f_+}{2\kappa}\mp\tfrac{\kappa}{2}$, and the dynamics reduces to an effective energy equation
\begin{equation}\label{energymain}
\frac{\dot R^2}{2}+V_{\rm eff}(R)=0,
\end{equation}
\begin{equation}\label{veffenergyeqmain}
V_{\rm eff}=\frac{c^2}{4}\left\{f_++f_-
-\left[\frac{(f_- - f_+)^2}{2\kappa^2}+\frac{\kappa^2}{2}\right]\right\},
\end{equation}
whose $\tau$-derivative gives the equation of motion ($'\equiv d/dR$)
\begin{equation}\label{equationofmotionmain}
\ddot R+V_{\rm eff}'(R)=0.
\end{equation}

Equations~\eqref{israeljunc0} have an equivalent canonical reading. Foliating by timelike $r=\mathrm{const}$ hypersurfaces, so that evolution runs along the normal rather than along time, the momentum conjugate to the induced metric $h_{ij}$ is $\Pi^{ij}=-\frac{c^{4}}{16\pi G}\sqrt{h}\,(\mathcal{K}^{ij}-\mathcal{K}h^{ij})$, and Eqs.~\eqref{israeljunc0} read $[h_{ij}]_\pm=0$, $[\Pi^{ij}]_\pm=-\tfrac12\sqrt{h}\,S^{ij}$: the configuration variable is continuous across $\Sigma$ while its conjugate momentum is not, the discontinuity being the surface stress\footnote{The second condition follows from \eqref{israeljunc0} on eliminating the trace, $[\mathcal{K}]_\pm=-(4\pi G/c^{4})S$. We keep the conventions of Eq.~\eqref{Kdef}; other sign and signature conventions for $\Pi^{ij}$ appear in the literature.}~\cite{Padmanabhan:2010zzb,Poisson:2009pwt}. The same $\Pi^{ij}$ is the boundary term in the variation of the gravitational action, from which the quasilocal energy inside radius $r$ is constructed~\cite{Brown:1992br,Hawking:1995fd}, and the matching becomes an energy balance at $R_0$: the energy just outside the shell exceeds that just inside by the shell's proper rest energy\footnote{For the metric~\eqref{metricSdSmain} the quasilocal energy is $E_\pm(r)=(c^{4}r/G)\,[\,1-\sqrt{f_\pm(r)}\,]$, the unit subtraction being the flat-space reference. At $\dot R=0$, Eq.~\eqref{betaeqm} then gives $E_+(R_0)-E_-(R_0)=(c^{4}R_0/G)(\sqrt{f_-}-\sqrt{f_+})=4\pi R_0^{2}\sigma_0c^{2}$, the reference term cancelling because both faces of $\Sigma$ share the same induced metric.} $4\pi R_0^{2}\sigma_0c^{2}$. Both energies are surface integrals on the same sphere, so this is a difference of boundary terms rather than a volume integral of an energy density. For $\Lambda_\pm=0$ this carries asymptotic information, since $E_+(r)\to m_+c^{2}$ as $r\to\infty$, the ADM energy~\cite{Arnowitt:1962hi} recovered as the Regge--Teitelboim boundary term at spatial infinity~\cite{Regge:1974zd}, so the matching relates the shell's content to a conserved charge read off far away. This is what makes $m_+$ an observable rather than a parameter of $f_+$. The balance then reads $m_+=m_-+4\pi R_0^{2}\sigma_0\sqrt{f_-(R_0)}-8\pi^{2}GR_0^{3}\sigma_0^{2}/c^{2}$: the asymptotic charge exceeds the interior mass by the shell's redshifted rest energy minus its binding energy.

For $\Lambda_\pm\neq0$ the static region ends at the cosmological horizon and no spatial infinity remains at which to evaluate $\Pi^{ij}$, so the masses are instead defined quasilocally~\cite{Misner:1964je,Hayward:1994bu}; the balance is unchanged, since it compares the two sides at the common radius\footnote{Conserved charges do exist for asymptotically de~Sitter spacetimes, but not as boundary terms on the timelike foliation used here: the Abbott--Deser construction~\cite{Abbott:1981ff} defines them relative to a background through its Killing vectors, while the covariant phase-space charges are defined at the spacelike conformal boundary~\cite{Ashtekar:2014zfa}. The Misner--Sharp mass for \eqref{frmetricmain} is $m^\pm_{\rm MS}(r)=m_\pm+c^{2}\Lambda_\pm r^{3}/(6G)$, so $m_\pm$ is recovered at any radius once the vacuum-energy contribution is subtracted. In the spherically symmetric canonical theory the mass is itself a canonical variable~\cite{Kuchar:1994zk,Hajicek:1997ry,Louko:1997hp}.} $R_0$.

 \begin{table*}
\caption{Stability windows at $\lambda=1$ (test-shell limit, App.~\ref{app:analytic}), assuming $9G^2m_\pm^2\Lambda_\pm/c^4\ll1$.
The branch is fixed by which term dominates $\Delta\equiv f_--f_+$: the mass term
($u\equiv2G \bar m/c^2R$, where $\bar m\equiv(m_++m_-)/2$) gives the photon-sphere window $0<w_0<\tfrac12$ for any
$\Lambda_\pm$ ordering; the cosmological term ($v\equiv\bar \Lambda R^2/3$, where $\bar \Lambda\equiv(\Lambda_++\Lambda_-)/2$) gives the
cosmological-horizon branch for $\Lambda_+=\Lambda_-$; where both are negligible
(static radius) the sign of $\delta\Lambda$ sets the window via
$\rho+1=\delta\Lambda\,R^2/\Delta$.}
\label{tab:windows}
\begin{ruledtabular}
\begin{tabular}{llll}
Scale of $R_0$ & Ordering & Dominant term ($\rho_0\equiv R_0\Delta'_0/\Delta_0$)  & Stable window \\
\hline
photon sphere & any & $u_0$ dominant, $\rho_0=-1$ & $0<w_0<\tfrac12$\footnote{The edge $w_0=\tfrac12$ corresponds to $R_0=3Gm_-/c^2$ (the photon sphere).} \\[2pt]
cosm.\ horizon & $\Lambda_+=\Lambda_-$ & $v_0$ dominant, $\rho_0=-1$ & $\dfrac{1-\sqrt{13}}{6}<w_0<0$ \\[6pt]
static radius & $\Lambda_+>\Lambda_-$ & $u_0,v_0\ll1$, $\rho_0=-1-2w_0$ & $-\tfrac23<w_0<0$ \\[2pt]
static radius & $\Lambda_+<\Lambda_-$ & $u_0,v_0\ll1$, $\rho_0=-1-2w_0$ & $w_0>0$\footnote{For $\Lambda_+<\Lambda_-$, $\delta\Lambda<0$ forces $w_0>0$ via
Eq.~\eqref{identitybounds}: there is no tension branch.Over $0<w_0<\tfrac12$ this
window overlaps the photon-sphere row, so at the same $(\lambda,w_0)$ the equilibrium
system admits two distinct solutions $(R_0,\sigma_0)$, one near the photon sphere
$3Gm_-/c^2$ and one at the SdS static radius $R_{\mathrm{st}}\equiv(3Gm_-/\bar \Lambda c^{2})^{1/3}$.} \\
\end{tabular}
\end{ruledtabular}
\end{table*}

\subsection{Linear equation of state}
\label{sec:eos}
 
The adiabatic sound speed of a surface layer with $p=p(\sigma)$ is $c_s^2=dp/d\sigma$. For $p=w\sigma c^2$ ($w$ constant), $c_s^2=wc^2$, so $w<0$ gives an imaginary sound speed and a gradient instability. This is avoided by a more general linear response: near equilibrium any barotropic $p(\sigma)$ admits a first-order Taylor expansion~\cite{Brady:1991np,Babichev:2004qp,Babichev:2004yx}, and we adopt
\begin{equation}\label{linearEoS}
p = \lambda (\sigma - \sigma_1) c^2,\qquad w(\sigma)\equiv \lambda\,\frac{\sigma-\sigma_1}{\sigma},
\end{equation} with constant slope $\lambda$ and reference density $\sigma_1$, so $c_s^2=\lambda c^2$ ($0\le\lambda\le1$). Substituting \eqref{linearEoS} into the conservation law \eqref{conservationofenergysigma1} and integrating gives the surface density \begin{equation}\label{solutionsigma}
\sigma(R) = \frac{\lambda \sigma_1}{1+\lambda} + C\, R^{-2(1+\lambda)}
= \frac{\lambda \sigma_1}{1+\lambda} + D\left(\frac{R_0}{R}\right)^{2(1+\lambda)},
\end{equation} 
with integration constant $C$ and $D\equiv C R_0^{-2(\lambda+1)}$. For the shell this recasts as
\begin{equation}\label{EqsigmaR}
\sigma(R) = \frac{\sigma_0}{1+\lambda}
\left[\lambda - w_{0}+ (1 + w_{0})\left(\frac{R_{0}}{R}\right)^{2(\lambda+1)}\right],
\end{equation}
where $\sigma_0\equiv\sigma(R_0)$ and $w_0\equiv p_0/(\sigma_0 c^2)$, so that $D=\tfrac{1+w_0}{1+\lambda}\sigma_0$ and $\sigma_1=\tfrac{\lambda-w_0}{\lambda}\sigma_0$ (equivalently $p=p_0+c_s^2(\sigma-\sigma_0)$ about equilibrium). Only the product $\lambda\sigma_1=(\lambda-w_0)\sigma_0$ enters
Eqs.~\eqref{EqsigmaR}--\eqref{pofR}, so the parametrisation is regular at
$\lambda=0$ even though $\sigma_1$ is not: there the layer has constant
pressure $p=w_0\sigma_0c^{2}$ and
$\sigma(R)=\sigma_0[(1+w_0)(R_0/R)^{2}-w_0]$. Equation~\eqref{EqsigmaR} holds wherever \eqref{linearEoS} is valid, while the stability analysis requires only that $p(\sigma)$ be linear in a neighborhood of $R_0$. The corresponding pressure is
\begin{equation}\label{pofR}
p(R)=\frac{c^2\sigma_0}{1+\lambda}
\left[(w_0-\lambda)+(1+w_0)\,\lambda\left(\frac{R_0}{R}\right)^{2(\lambda+1)}\right].
\end{equation}
For $\lambda=w_0$ ($\sigma_1=0$) we recover the standard law $p=w_0\sigma c^2$ at $R_0$, with the familiar form~\cite{Alestas:2020wwa}
\begin{equation}\label{sigmaperivpaper}
\sigma(R) = \sigma_0 \left(\frac{R_0}{R}\right)^{2(1+w_0)}.
\end{equation}

Expanding \eqref{EqsigmaR} to first order about the equilibrium radius $R_0$, with $\delta\sigma\equiv\sigma-\sigma_0$ and $\delta p\equiv p-w_0\sigma_0 c^2$,
\begin{equation}\label{deltasdp}
\delta\sigma = -\frac{2(1+w_0)\,\sigma_0\,\delta R}{R_0},\qquad
\delta p = -\frac{2\lambda(1+w_0)\,\sigma_0\,c^2\,\delta R}{R_0}.
\end{equation}
For $-1<w_0<0$ an outward perturbation lowers $\sigma$ and makes $p$ more negative (more tension), pulling the shell back, while an inward one raises $\sigma$ and weakens the tension. Positive-pressure shells ($w_0>0$) behave analogously: contraction raises both $\sigma$ and $p$, expansion lowers them. Since $w(\sigma)$ can change sign as $\sigma(R)$ varies about $R_0$ (for $\lambda>0$), stability combines both: a tension shell must sit in repulsive, or sufficiently weakly attractive, gravity, and a positive-pressure shell in attractive, or sufficiently weakly repulsive, gravity. By contrast, phantom matter ($w_0<-1$) amplifies perturbations and is destabilizing for $\lambda>0$; for $\lambda=w_0$, $\delta p\sim w_0(1+w_0)$, which for $w_0<0$ interchanges the stabilizing and destabilizing responses between $-1<w_0<0$ and $w_0<-1$.

Both $\delta\sigma$ and $\delta p$ in Eq.~\eqref{deltasdp} are proportional
to $(1+w_0)$ and vanish at $w_0=-1$, where Eq.~\eqref{EqsigmaR} gives constant $\sigma$: the
layer has fixed mass per unit area and its surface stress does not respond to radial
displacement. Two familiar objects are described by $w_0=-1$. The first is the wall of a
thin-wall vacuum-decay bubble
\cite{Zeldovich:1974uw,Kibble:1976sj,Coleman:1977py,Callan:1977pt,Coleman:1980aw,Vilenkin:1984ib},
made of a scalar field that separates a false vacuum from a true one. Its tension and the
vacuum-energy jump across it both follow from the scalar potential, and $p=-\sigma c^{2}$
holds only when the wall is thin compared with its radius. The second is the
Brown--Teitelboim membrane \cite{Teitelboim:1985ya,Brown:1987dn,Brown:1988kg}, a fundamental
object charged under a three-form potential. Its tension is a free parameter, $\delta\Lambda$
is set by the charge, and $p=-\sigma c^{2}$ is exact. Both nucleate at rest and then expand,
so the radius assigned to them satisfies $\dot R=0$, which by Eq.~\eqref{energymain} means
$V_{\rm eff}=0$; for the membrane this is Eq.~\eqref{betaeqm} at $m_\pm=0$, that is in pure
vacuum with no central mass. A static shell must satisfy $V'_{\rm eff}=0$ as well. A
$w_0=-1$ layer can satisfy both conditions, but always with $V''_{\rm eff}<0$, since
$w_0=-1$ lies below every window of Sec.~\ref{sec:numerical}.

Equation~\eqref{EqsigmaR} splits into two barotropic components, since
$p=w\sigma c^{2}$ implies $\sigma\propto R^{-2(1+w)}$. The constant term is a $w=-1$ wall of
surface density $\sigma_{\rm DW}=(\lambda-w_0)\sigma_0/(1+\lambda)$ and pressure
$p_{\rm DW}=-\sigma_{\rm DW}c^{2}$; the remainder is a fluid with $w=\lambda$,
$\sigma_{\rm DF}=(1+w_0)\sigma_0(R_0/R)^{2(1+\lambda)}/(1+\lambda)$ and
$p_{\rm DF}=\lambda\sigma_{\rm DF}c^{2}$, both surface densities positive for
$-1<w_0<\lambda$. The combination $\sigma+p/c^{2}$ vanishes for the wall, so
Eq.~\eqref{conservationofenergysigma1} splits with no source term: $\sigma_{\rm DW}$ stays
constant and the fluid conserves its energy alone. The two are nonetheless coupled
gravitationally, since $V_{\rm eff}$ depends on $\kappa\propto\sigma_{\rm DW}+\sigma_{\rm
DF}$; what is fixed is only that the wall's stress does not respond to radial displacement,
so $\delta\sigma$ and $\delta p$ in Eq.~\eqref{deltasdp} come entirely from the fluid, which
is therefore what moves $w_0$ into the stable windows, and $c_s^{2}=\lambda c^{2}$ is
independent of $w_0$. The tension shells found below have $\Lambda_+\ge\Lambda_-$, and for
$\Lambda_+>\Lambda_-$ they may be read as a Brown--Teitelboim membrane carrying dark fluid,
this being the ordering favoured by nucleation. We do not impose the Brown--Teitelboim
relation between membrane charge, tension and $\delta\Lambda$, which here remain
independent; the reading is an analogy. We treat the layer phenomenologically, in terms of
$(\sigma_0,w_0,\lambda)$; none of the results depend on this reading.

\subsection{Equilibrium and stability conditions}\label{sec:equilibrium}

A static shell ($\dot R=\ddot R=0$) is an equilibrium configuration, $V_{\rm eff}(R_0)=V_{\rm eff}'(R_0)=0$, and is \emph{stable}~\cite{Visser:2003ge} against radial perturbations if $V_{\rm eff}''(R_0)>0$.
In the \emph{test-shell} limit $m_+/m_-\to1$, $\Lambda_+/\Lambda_-\to1$, the stability boundary admits an approximate analytic treatment, developed in full in Appendix~\ref{app:analytic}. With $\Delta\equiv f_--f_+$ and $g\equiv\Delta/2\kappa=O(1)$, the reduced potential is $V_{\rm eff}=\tfrac{c^2}{2}[f-g^2]+O(e^2)$, and the stable windows are fixed solely by whether the mass or $\Lambda$ term dominates $\Delta$ (i.e.\ by $\Lambda_+/\Lambda_-$). Throughout $f_0=O(1)$, since the stable radius sits well away from both horizons. At $\lambda=1$ this gives three regimes (see also Table \ref{tab:windows}):
\begin{itemize}
\item \emph{$R_0$ deep inside the static-radius scale} (mass-dominated $\Delta$): $V_{\rm eff,0}''\propto(1-2w_0)(3w_0+2)$, stable for $0<w_0<\tfrac12$, approaching, but never reaching, the photon sphere $R_0=3Gm_-/c^2$ as $w_0\to\tfrac12$.
\item \emph{$R_0$ far outside the static-radius scale, $\Lambda_+=\Lambda_-$}: $V_{\rm eff,0}''\propto-(3w_0^2-w_0-1)$, stable for $(1-\sqrt{13})/6<w_0<0$, the shell migrating toward---but never reaching---the cosmic horizon as $w_0\to(1-\sqrt{13})/6$.

\item \emph{Static-radius scale, $\Lambda_+\gtrless\Lambda_-$}: $V_{\rm eff,0}''\propto(3w_0+2)$, stable for $-\tfrac23<w_0<0$ if $\Lambda_+>\Lambda_-$, and for $w_0>0$ if $\Lambda_+<\Lambda_-$.
\end{itemize}

\subsection{Numerical results}
\label{sec:numerical}
 
\begin{figure*}
    \centering
    \includegraphics[width=1\linewidth, height=0.95\textheight, keepaspectratio]{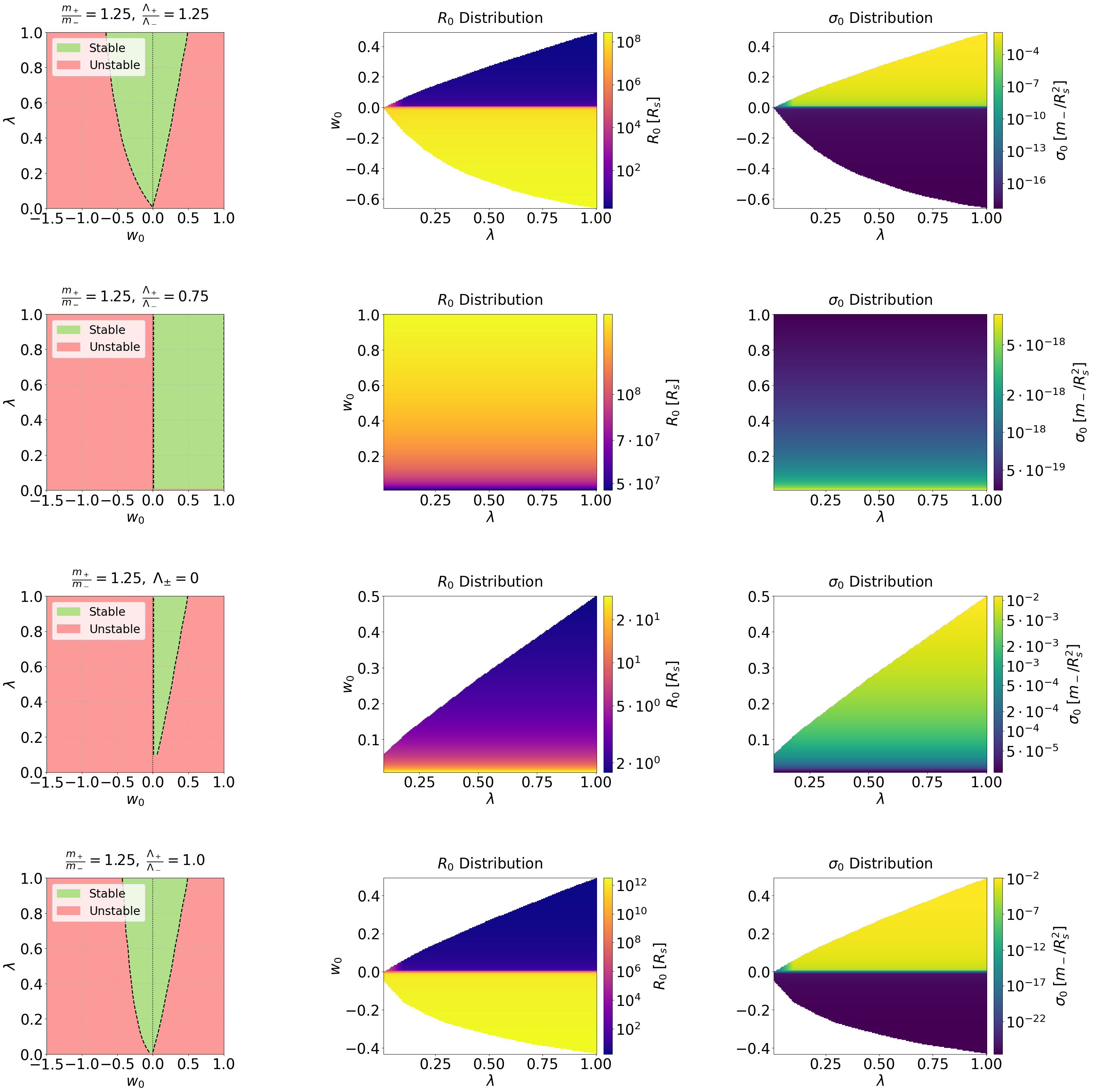}
   \caption{Stability and equilibrium of shells in the $(w_0,\lambda)$ parameter space, with $m_-=10^{10}\,M_\odot$ and $\rho_{\Lambda_+}=\rho_{\rm crit}=3H_0^2/(8\pi G)$~\cite{Planck2018}. We solve Eqs.~\eqref{systemA}--\eqref{systemB} for $(R_0,\sigma_0)$, requiring $V''_{\rm eff}(R_0)>0$ [Eq.~\eqref{Veffpp0}] and $\sigma_0>0$ [Eq.~\eqref{sigma0cond}].  \textbf{Left:} stable (green) versus unstable (red) regions. \textbf{Middle/Right:} the equilibrium radius $R_0$ (in units of $R_s\equiv 2 G m_-/c^2$) and surface density $\sigma_0$ (in units of $m_-/R_s^2$) over the stable domain. Rows: (a) $m_+/m_-=\Lambda_+/\Lambda_-=1.25$; (b) $m_+/m_-=1.25,\ \Lambda_+/\Lambda_-=0.75$; (c) $m_+/m_-=1.25,\ \Lambda_\pm=0$; (d) $m_+/m_-=1.25,\ \Lambda_+/\Lambda_-=1.0$. In row (b) stable shells exist for $0\lesssim w_0$. Over
$0\lesssim w_0\lesssim1/2$ there are two solutions $(R_0,\sigma_0)$ at the same
$(\lambda,w_0)$: separate configurations with different $\sigma_0$, not two minima of
one $V_{\rm eff}$. The middle and right panels show only the static-radius one; the other
sits at the photon-sphere scale, where the mass term dominates $\Delta$, so it is insensitive to $\Lambda_+/\Lambda_-$ and essentially coincides
with rows (a), (c) and (d) at the same $w_0$.
    Reddish or white areas mark where no static equilibrium with $\sigma_0>0$ exists,
or where it exists but is unstable ($V''_{\rm eff}(R_0)<0$).}
    \label{fig:combined_analysis_5cases}
\end{figure*}
\begin{figure*}
    \centering
    \includegraphics[width=\textwidth]{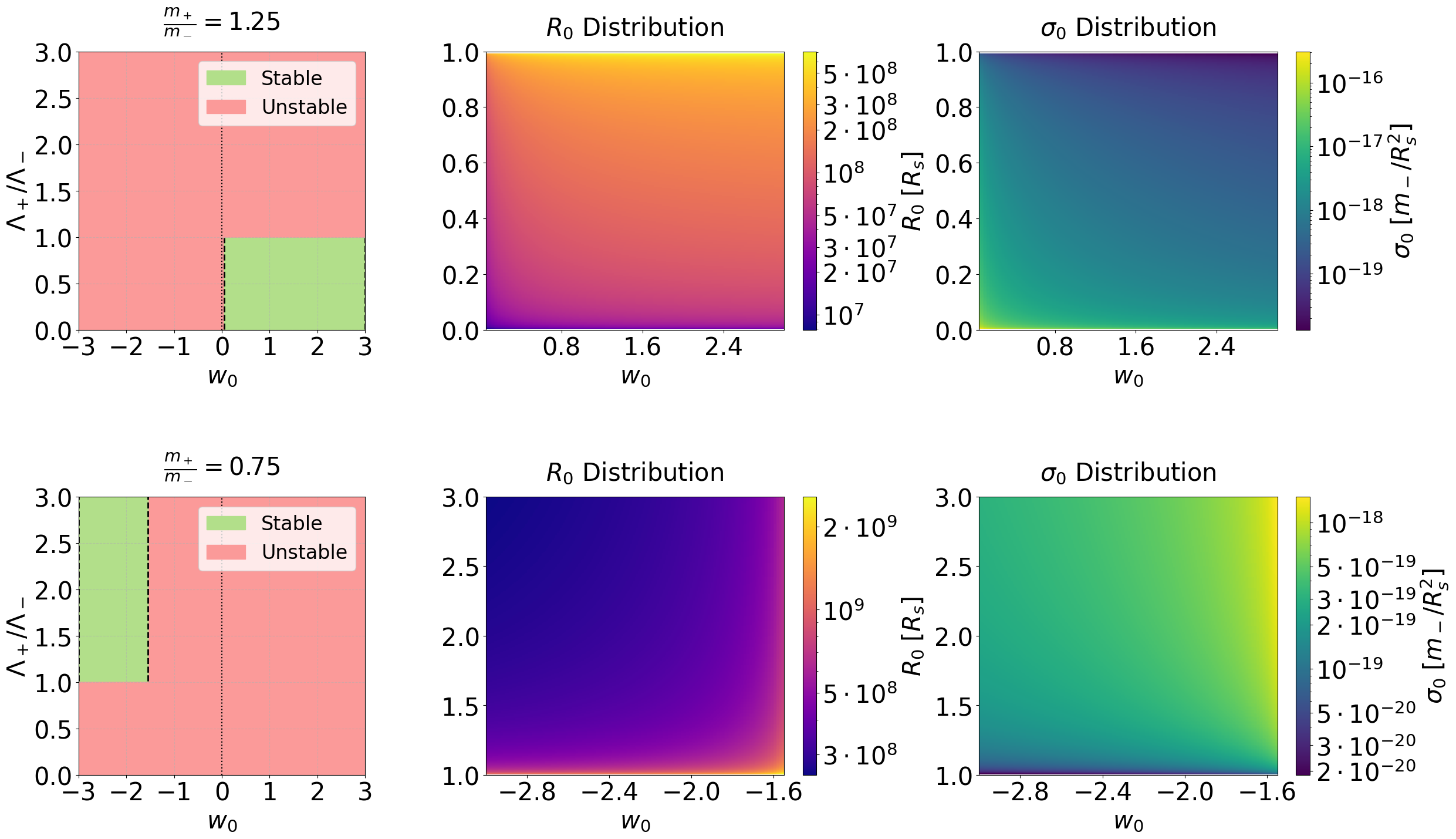}
    \caption{Stability and equilibrium in the $(w_0,\Lambda_+/\Lambda_-)$ parameter space for the case $w_0=\lambda$ [Eq.~\eqref{sigmaperivpaper}], with $m_-=10^{10}\,M_\odot$ and $\rho_{\Lambda_+}=\rho_{\rm crit}=3H_0^2/(8\pi G)$~\cite{Planck2018}. We solve  Eqs.~\eqref{systemA}--\eqref{systemB} for $(R_0,\sigma_0)$, requiring $V''_{\rm eff}(R_0)>0$  [Eq.~\eqref{Veffpp0}] and $\sigma_0>0$ [Eq.~\eqref{sigma0cond}]. \textbf{Left:} stable (green) versus unstable (red) regions. \textbf{Middle/Right:} $R_0$ (in units of $R_s$) and $\sigma_0$ (in units of $m_-/R_s^2$) over the stable domain. Top row $m_+/m_-=1.25$, bottom row $m_+/m_-=0.75$.}
    \label{fig:stability_analysisw0lam}
\end{figure*}

 Substituting \eqref{frmetricmain}, \eqref{kapparmain}, and \eqref{EqsigmaR} into \eqref{veffenergyeqmain} gives the explicit $V_{\rm eff}(R)$, Eq.~\eqref{veff_full} of Appendix~\ref{app:explicit}.  The full equilibrium system [Eqs.~\eqref{systemA}--\eqref{systemB}] and the explicit form of $V_{\rm eff}''(R_0)$ [Eq.~\eqref{Veffpp0}] are given in Appendix~\ref{app:explicit}. We require $\sigma_0>0$, i.e.
\begin{equation}\label{sigma0cond}
-\frac{2G}{c^2 R_0} [m]_{\pm} < \frac{R_0^2}{3} [\Lambda]_{\pm}.
\end{equation}
We identify the viable equilibria $(R_0,\sigma_0)$ for a given $(m_\pm,\Lambda_\pm,\lambda,w_0)$ by solving Eqs.~\eqref{systemA}--\eqref{systemB} and retaining those with $V''_{\rm eff}(R_0)>0$ [Eq.~\eqref{Veffpp0}] and $\sigma_0>0$ [Eq.~\eqref{sigma0cond}]. A shell lies in the static region common to both Schwarzschild--de Sitter patches, i.e.\ $f_\pm(R_0)>0$, above the interior black-hole horizon and below the exterior cosmological horizon. Such a region exists, and can be occupied by the shell, only if both spacetimes satisfy the Nariai bound, $9G^2m_\pm^2\Lambda_\pm/c^4<1$. Our results  assume that $9G^2m_\pm^2\Lambda_\pm/c^4\ll1$, a condition satisfied by all astrophysical black holes.

As shown in Figs.~\ref{fig:combined_analysis_5cases}--\ref{fig:stability_analysisw0lam}, both the existence and the size of the stability region\footnote{Although the stable region and the characteristic scales of $R_0$ and $\sigma_0$ depend on the  $9G^2m_\pm^2\Lambda_\pm/c^4$ and shrink as it approaches unity, this is immaterial here: even for $m_-=10^{10}\,M_\odot$, near the upper end of the observed black-hole mass range~\cite{Natarajan:2008ks}, it is only $\sim10^{-25}$ at the Planck~2018 $\Lambda$, so Figs.~\ref{fig:combined_analysis_5cases}--\ref{fig:stability_analysisw0lam} are unaffected.} in the $w_0$--$\lambda$ plane (for $\sigma_0>0$) depend sensitively on the mass and cosmological-constant ratios. Fixing $m_-=10^{10}\,M_\odot$ and $\rho_{\Lambda_+}=\rho_{\rm crit}$~\cite{Planck2018}, we find that static stable thin shells occur only in\footnote{Static stable configurations with $m_+/m_-<1$ exist but require $\lambda<0$, leading to gradient instability~\cite{Babichev:2004yx}.} $\frac{m_+}{m_-}>1$ with $\frac{\Lambda_+}{\Lambda_-}>1$, $=1$, or $<1$, plus the matched Schwarzschild case $\frac{m_+}{m_-}>1,\ \Lambda_+=\Lambda_-=0$. The overall shape of the stable region depends only on whether the ratios exceed, equal, or fall below unity; their precise values instead set $\sigma_0$ and $R_0$ (Fig.~\ref{fig:combined_analysis_5cases}). The case $\lambda=1$ yields the widest allowed range of $w_0$. 

Setting $\lambda=w_0$, equivalently $\sigma_1=0$, reduces the equation of state to the single-parameter law $p=w_0\sigma c^2$ [Eq.~\eqref{sigmaperivpaper}], for which $c_s^2=w_0c^2$. This case yields stable configurations approximately for $0\lesssim w_0$ when $m_+/m_->1$, $\Lambda_+/\Lambda_-<1$, and for $w_0\lesssim-1.5$ when $m_+/m_-<1$, $\Lambda_+/\Lambda_->1$ (Fig.~\ref{fig:stability_analysisw0lam}); stability is always lost at $\lambda=w_0=0$. The $0\lesssim w_0$ branch ($\Lambda_+/\Lambda_-<1$, radii near the static-radius scale) has real $c_s$ and is already contained in the broader general $(\lambda,w_0)$ analysis, adding no new qualitative features. The $w_0\lesssim-1.5$ branch has $\lambda=w_0<0$, hence $c_s^2<0$, and is excluded by the resulting gradient instability~\cite{Babichev:2004yx}.

For $\Lambda_+>\Lambda_-$ ($-2/3\lesssim w_0\lesssim1/2$), positive-pressure shells sit near the photon sphere while negative-pressure shells (appearing smoothly at $w_0=0$) reach only the static-radius scale. For $\Lambda_+<\Lambda_-$ the static stable shells lie either at the static-radius scale, for $w_0\gtrsim0$, or at the photon-sphere scale, for $0\lesssim w_0\lesssim1/2$. Over $0\lesssim w_0\lesssim1/2$ the equilibrium system therefore has two distinct stable roots at the same $(\lambda,w_0)$: the photon-sphere solution approaches the photon sphere as $w_0\to1/2$, while the static-radius solution moves outward as $w_0$ increases. The two carry different $\sigma_0$ and hence different $\sigma(R)$, so each is a separate configuration with its own effective potential; they are not two minima of one $V_{\rm eff}$ (see also Appendix~\ref{app:explicit}). In the case $\Lambda_+=\Lambda_-\equiv\Lambda$ ($\rho_\Lambda=\rho_{\rm crit}$~\cite{Planck2018}), where the two SdS regions differ only in mass, $\sigma_0>0$ in \eqref{betaeqm} requires $[m]_\pm>0$, and stable solutions span $(1-\sqrt{13})/6\lesssim w_0\lesssim\tfrac12$ (Fig.~\ref{fig:combined_analysis_5cases}), with positive-pressure shells near the photon sphere and decreasing $w_0<0$ driving them toward the cosmological horizon. \emph{Overall, static stable tension shells ($\sigma_0>0$, $w_0<0$, $\lambda>0$) exist only for $m_+/m_->1$ and $\Lambda_+/\Lambda_-\ge1$.}

 \begin{figure*}
    \centering
    \begin{subfigure}[b]{0.45\textwidth}\centering
        \includegraphics[width=\textwidth]{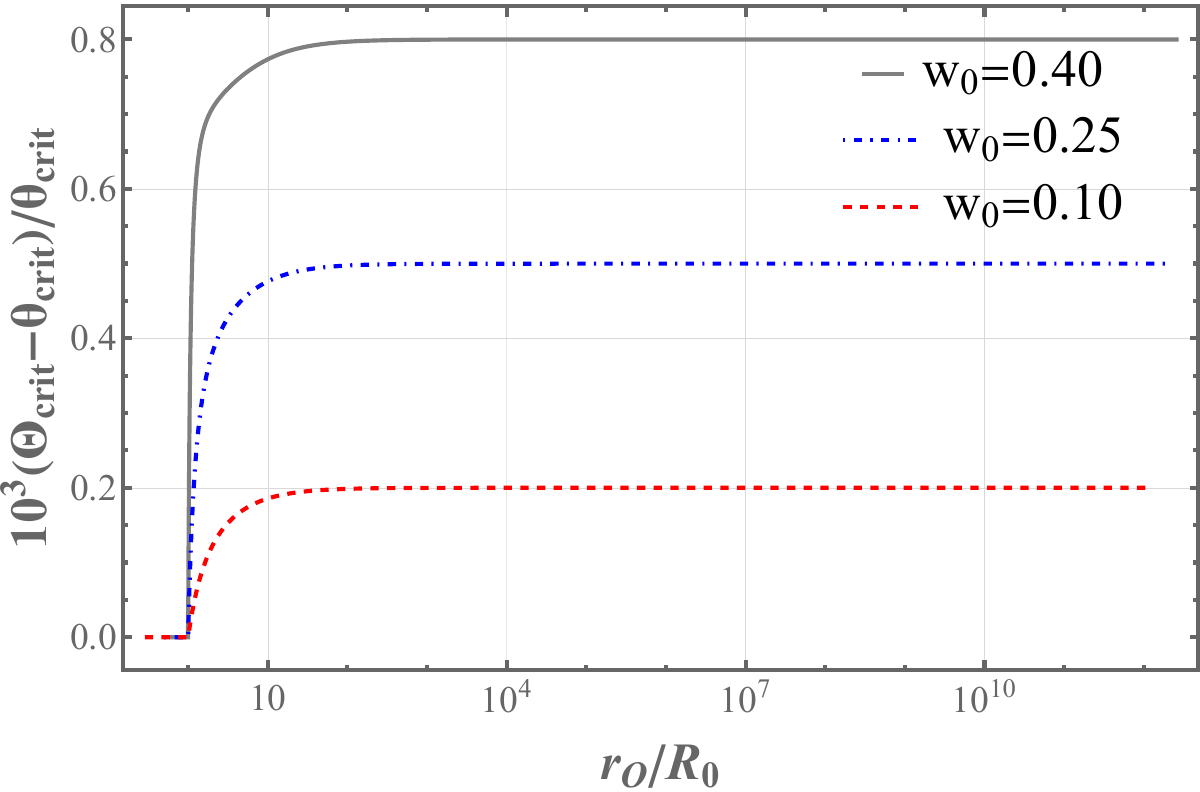}\label{fig:shmod1}\end{subfigure}
    \begin{subfigure}[b]{0.45\textwidth}\centering
        \includegraphics[width=\textwidth]{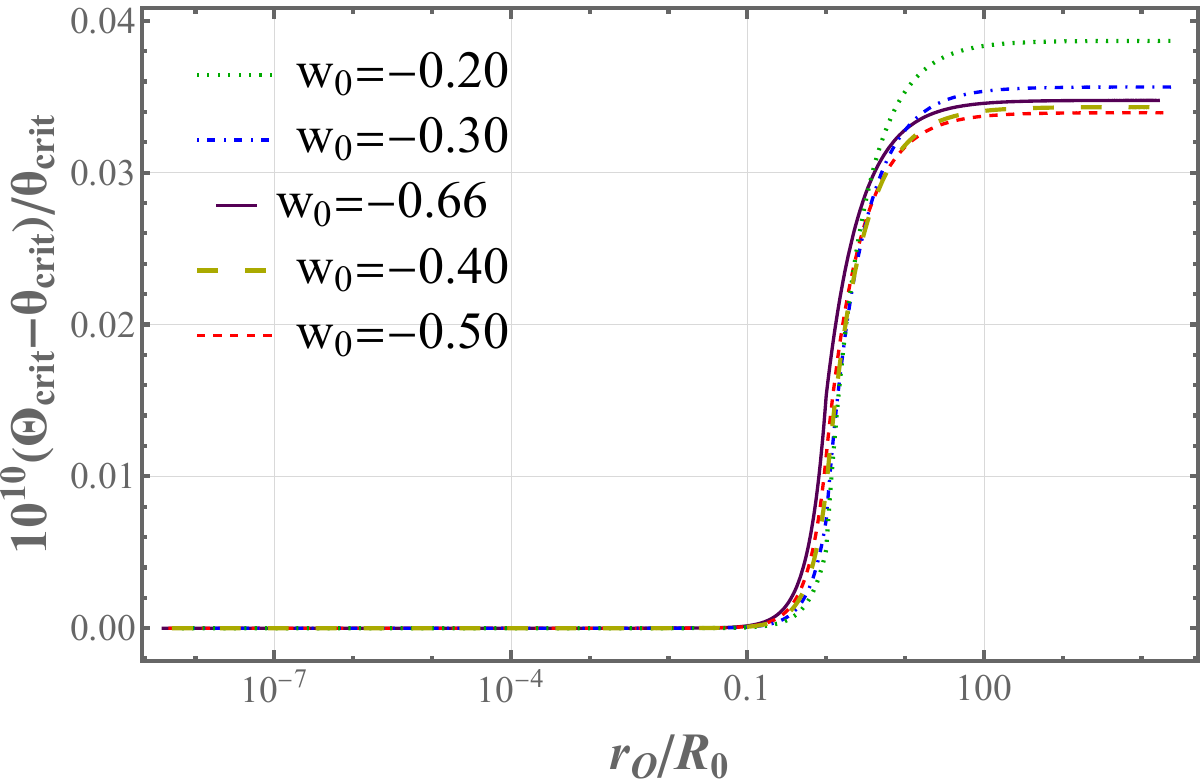}\label{fig:shmod2}\end{subfigure}
    \begin{subfigure}[b]{0.45\textwidth}\centering
        \includegraphics[width=\textwidth]{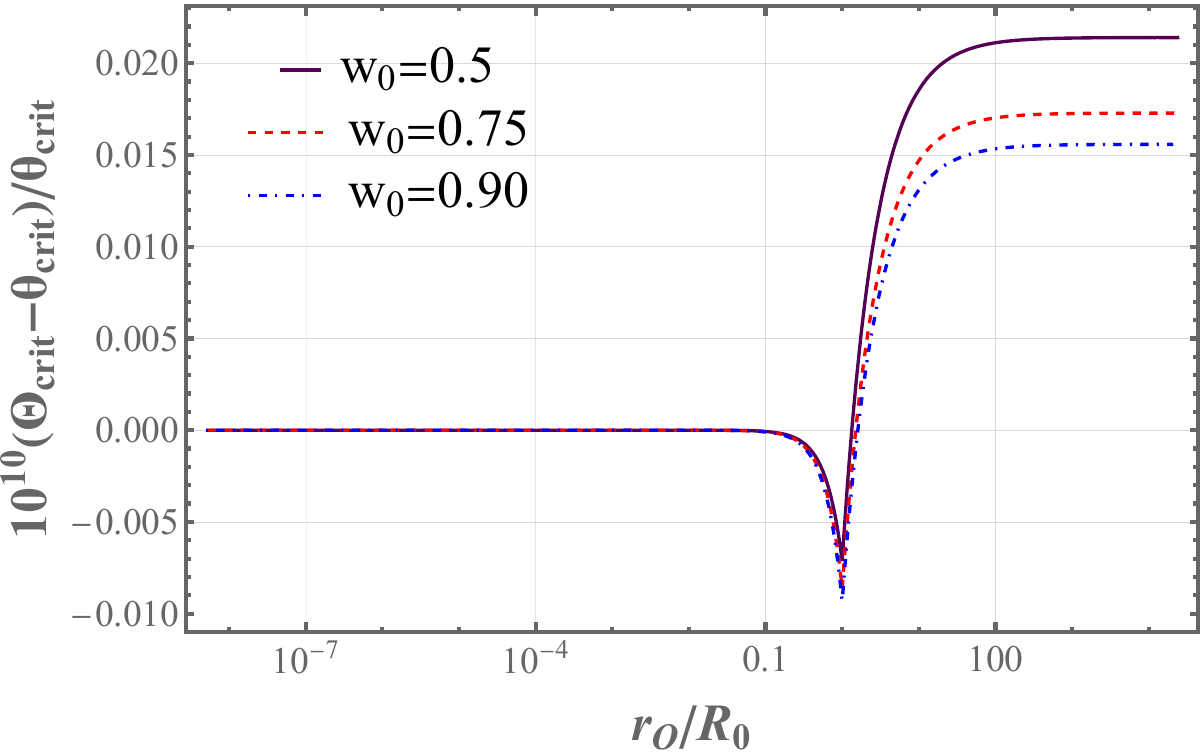}\label{fig:shmod3}\end{subfigure}
    \caption{Relative shadow deviation $(\Theta_{\rm crit}-\theta_{\rm crit})/\theta_{\rm crit}$ versus $r_O/R_0$, from Eqs.~\eqref{sdsthetacrit} and \eqref{Thetactita}, for $\lambda=1$, $m_+/m_-=1.001$, and $\rho_{\Lambda_+}=\rho_{\rm crit}$~\cite{Planck2018}. \textbf{(a)} Positive-pressure shells near the photon sphere ($\Lambda_+/\Lambda_-=1.001$); deviations reach $\sim10^{-3}$, growing as larger $w_0$ pushes the shell toward the photon sphere ($10^3$ scaling). \textbf{(b)} Negative-pressure shells near the static radius ($\Lambda_+/\Lambda_-=1.001$; $10^{10}$ scaling), much smaller. \textbf{(c)} Positive-pressure shells near the static radius ($\Lambda_+/\Lambda_-=0.999$; $10^{10}$ scaling), comparable to (b). For the negative-pressure shells the deviation decreases as $w_0$ drops from $0$ to $-1/2$, minimizes at $w_0=-1/2$, and rises again as $w_0\to-2/3$.}
    \label{fig:stab}
\end{figure*}

\section{Observational Signatures of a Static Stable Dark Fluid Shell}
\label{sec:signatures}
 
The observational signatures of thin-shell and shell-like structures in black-hole, gravastar, and
wormhole spacetimes---on strong-field lensing, shadow morphology, and photon rings---have been
studied widely~\cite{Sakai:2014pga,Antoniou:2022dre,Peng:2021osd,He:2024yeg,Cao:2025qzy,Saleem:2025axo,Macedo:2025ipc,Laeuger:2025zgb,Wu:2024juj,BenAchour:2025uzp,Konoplya:2025mvj}.
We model the shell as a transparent dark fluid: since it does not couple to the electromagnetic
field, it neither emits nor absorbs radiation and cannot be observed directly; it interacts only
gravitationally, however, so it deflects light traversing it and can, at least in principle, imprint
on the black-hole shadow~\cite{Cao:2025qzy}.

For a pure SdS spacetime (no shell, $m=m_-$, Planck $\Lambda$~\cite{Planck2018}), the
static-observer shadow angular radius is (Appendix~\ref{sec:appendixa})~\cite{PhysRevD.60.044006}
\begin{equation}\label{sdsthetacrit}
\sin^2\theta_{\rm crit}=\frac{f(r_O)\,b_{\rm crit}^{\,2}}{r_O^2},
\end{equation}
with $r_O$ the observer radius. Fixing $\lambda=1$ (which sets the $w_0$ range without affecting
the deviation magnitude), $m_+/m_-=1.001$ and $\Lambda_+/\Lambda_-=1.001$ or $0.999$, $m_-=10^{10}M_\odot$, and
$\rho_{\Lambda_+}=\rho_{\rm crit}$~\cite{Planck2018}, we compute the shadow for a static observer
at $r_O$ and compare to~\eqref{sdsthetacrit}. In the equatorial plane the null geodesics obey
\begin{align}
\frac{dt_\pm}{d\zeta}&=\frac{k_\pm}{f_\pm(r_\pm)},\label{p0eq}\\
\frac{d\phi}{d\zeta}&=\frac{h_\pm}{r_\pm^2},\label{dotphi}\\
c^2k_\pm^2-\frac{h_\pm^2}{r_\pm^2}f_\pm(r_\pm)&=\left(\frac{dr_\pm}{d\zeta}\right)^2,\label{G3}
\end{align}
with $k_\pm,h_\pm$ constants. For an observer inside the shell ($r_-=r_O<R_0$) with the shell
outside the inner photon orbit ($R_0>3Gm_-/c^2$), an orthonormal tetrad
(Appendix~\ref{sec:appendixa})~\cite{PhysRevD.60.044006,Perlick:2018iye} gives
\begin{equation}\label{eq:theta-finala}
\sin^2\theta^-_{\rm crit}=\frac{f_-(r_O)\,(b_{\rm crit}^{-})^2}{r_O^2},
\end{equation}
where for the circular photon orbit $r_-=3Gm_-/c^2$ [Eq.~(\ref{orbiteq})]
\begin{equation}\label{criticalimpactparameter}
b^-_{\rm crit}=\frac{3\sqrt{3}\,Gm_-}{c^2\sqrt{1-9G^2m_-^2\Lambda_-/c^4}}.
\end{equation}
For a static shell ($\dot R=0$), the metric~\eqref{metricSdSmain} at $r_\pm=R$ gives
$dt_+/dt_-=\sqrt{f_-(R)/f_+(R)}$, and continuity of $\phi$ gives $h_+=h_-=h$, so
\begin{equation}\label{bratio}
\frac{b_+}{b_-}=\frac{k_-}{k_+}=\sqrt{\frac{f_-(R)}{f_+(R)}}.
\end{equation}
For an observer at the shell ($r_O=R$), $\sin^2\theta^{\pm}_{R,\rm crit}=f_-(R)(b^-_{\rm crit})^2/R^2
=f_+(R)(b^+_{\rm crit})^2/R^2$, and for $r_+=r_O>R$, $\sin^2\theta^+_{\rm crit}
=\frac{f_+(r_O)R^2}{f_+(R)r_O^2}\sin^2\theta^{\pm}_{R,\rm crit}$. Combining, the shadow seen at
arbitrary $r_O$ relative to a static shell of radius $R_0$ is\footnote{The exterior region carries its own unstable circular photon
orbit at $r=3Gm_+/c^{2}$, which lies in $\mathcal{M}^{+}$ whenever
$R_0<3Gm_+/c^{2}$. With two barriers the shadow edge is set by the smaller
critical impact parameter, since a ray is captured only if it clears both. The
two are defined by different conserved energies $k_\pm$, so they cannot be
compared directly: by \eqref{bratio}, the inner critical parameter measured by
an outside observer is $b^{-}_{\rm crit}\sqrt{f_-(R_0)/f_+(R_0)}$, and
\eqref{Thetactita} holds provided this is smaller than $b^{+}_{\rm crit}$.
Neglecting $\Lambda_\pm$ at these radii, the condition reads
$m_-^{2}f_-(R_0)<m_+^{2}f_+(R_0)$, i.e.
$R_0>\frac{2Gm_-}{c^{2}}\frac{1+\mu+\mu^{2}}{1+\mu}$ with $\mu\equiv m_+/m_-$,
reducing to the photon sphere $3Gm_-/c^{2}$ as $\mu\to1$. It holds throughout
Fig.~\ref{fig:stab}(a); were it violated, the exterior orbit would set the
shadow and add a second photon ring.}
\begin{equation}\label{Thetactita}
\sin^2\Theta_{\rm crit}=
\begin{cases}
\dfrac{f_-(r_O)\,(b_{\rm crit}^{-})^2}{r_O^2} & r_O\le R_0,\\[2mm]
\dfrac{f_+(r_O)\,R_0^2}{f_+(R_0)\,r_O^2}\,\sin^2\theta^{\pm}_{R_0,\rm crit} & r_O>R_0.
\end{cases}
\end{equation}

Both signs of $w_0$ produce effects, mainly for observers outside the shell. Throughout we fix the exterior value $\rho_{\Lambda_+}=\rho_{\rm crit}$~\cite{Planck2018} and realize the orderings $\Lambda_+/\Lambda_-\gtrless1$ by varying the interior $\Lambda_-$. For $w_0<0$ the relative deviation $(\Theta_{\rm crit}-\theta_{\rm crit})/\theta_{\rm crit}$ is tiny ($\sim10^{-12}$) at fixed $(m_\pm,\Lambda_\pm)$, since these shells sit near the static radius (Fig.~\ref{fig:stab}); near-static-radius positive-pressure shells with $\Lambda_+/\Lambda_-<1$ give the same order with distinct behavior. For $w_0>0$ and an observer near the black hole the deviation reaches $\sim10^{-3}$ even for $m_+/m_-=1.001$. Relative to pure SdS at $\rho_\Lambda=\rho_{\rm crit}$~\cite{Planck2018}: positive-pressure shells close to the black hole vary steeply [Fig.~\ref{fig:stab}(a)]; for negative-pressure shells with $\Lambda_+>\Lambda_-$ ($\Lambda_-<\Lambda<\Lambda_+$) the shadow grows smoothly for observers inside $R_0$ and rises even before reaching the shell\footnote{The drift, when  $r_{O}<R_0$ in Figs.~3(b)--(c), reflects our choice of
reference rather than the shell. We compare to pure SdS with $(m_-,\Lambda)$ at
$\rho_\Lambda=\rho_{\rm crit}=\rho_{\Lambda_+}$: outside the shell the
two spacetimes then share $\Lambda$ and differ only through $m_+$ vs $m_-$,
whereas inside they
share $m_-$ and differ only through $\Lambda_-$ vs $\Lambda_+$. The interior deviation is
therefore just this $\delta\Lambda$ mismatch, visible only where $\Lambda_-r_O^2/3$ is comparable
to $2Gm_-/(c^2r_O)$, i.e. at the static-radius scale, and with the sign of the ordering: positive
for $\Lambda_+>\Lambda_-$, so it runs into the step at $R_0$ as one smooth rise (b), and negative
for $\Lambda_+<\Lambda_-$, so the curve first falls and is then carried back up through zero to a
positive plateau (c). On the photon-sphere branch $R_0\ll R_{\rm st}$ and the $\Lambda$ term is
negligible over the whole interior, so the deviation vanishes for $r_O<R_0$ and the crossing is a
sharp step (a). } [Fig.~\ref{fig:stab}(b)]; for $\Lambda_+<\Lambda_-$ ($\Lambda_->\Lambda$) it first shrinks inside $R_0$ then grows beyond, matching (b) in amplitude but differing in detail [Fig.~\ref{fig:stab}(c)]. For the negative-pressure shells of Fig.~\ref{fig:stab}(b), seen by distant observers, $\Theta_{\rm crit}-\theta_{\rm crit}$ decreases as $w_0$ drops from $0$ to $-\tfrac12$,
minimizes at $w_0=-\tfrac12$, and rises again as $w_0\to-\tfrac23$.

\section{Conclusion}
\label{sec:conclusion}

We analyzed the existence and radial stability of static, spherically symmetric \emph{dark fluid} thin shells joining two SdS spacetimes $(m_\pm,\Lambda_\pm)$---transparent layers coupling to photons only gravitationally. With the Israel formalism and the linearized law $p=p_0+c_s^2(\sigma-\sigma_0)$, whose slope $c_s^2=\lambda c^2$ is independent of $w_0\equiv p_0/(\sigma_0 c^2)$, we found radially stable shells of both pressure signs, notably tension shells ($w_0<0$) with real $c_s$.

\emph{Where can such a layer sit stably?} Only for $m_+/m_->1$, at three characteristic scales: the photon sphere, the SdS static radius, and the cosmological horizon. We systematically map the stable equilibria numerically over $(m_\pm,\Lambda_\pm,\lambda,w_0)$ and confirm the window edges with an approximate analytic treatment in the test-shell limit (Appendix~\ref{app:analytic}). At $\lambda=1$:
\begin{itemize}
    \item $\Lambda_+=\Lambda_-$: $(1-\sqrt{13})/6\lesssim w_0\lesssim 1/2$. As $w_0$ decreases the shell moves outward from the photon sphere, and for $w_0<0$ toward the cosmological horizon.
    \item $\Lambda_+>\Lambda_-$: $-2/3\lesssim w_0\lesssim 1/2$. As $w_0$ turns negative the shell moves out  to the static-radius scale.
        \item $\Lambda_+<\Lambda_-$: $0\lesssim w_0$. For $w_0\gtrsim0$ the shell sits at the static-radius scale and moves outward as $w_0$ grows; over $0\lesssim w_0\lesssim1/2$ a second, distinct stable solution exists near the photon sphere.
\end{itemize}
For $0\lesssim w_0\lesssim1/2$ all three orderings share a common photon-sphere branch: the shell lies outside the photon sphere, approaching but never reaching it, where the mass term dominates and $\Lambda_\pm$ is negligible.  This reduces to the pure-Schwarzschild result of Brady--Louko--Poisson~\cite{Brady:1991np}, whose photon-sphere bound is the upper edge $w_0=1/2$; the orderings differ only outside this range, where the cosmological constant matters; the windows depend only on whether the ratios exceed, equal, or fall below unity, while their values fix $R_0$ and $\sigma_0$. Tension shells\footnote{Evaluated on the shell's surface stress--energy, the energy conditions \cite{Visser:2003ge} reduce to inequalities on $w_0$ at $R_0$, checked over the stable parameter space. With $\sigma_0>0$, the null and weak energy conditions hold for $w_0\ge-1$ and are thus satisfied across all windows. The dominant energy condition holds for $|w_0|\le1$, i.e.\ provided $0\lesssim w_0\le1$ on the $\Lambda_+<\Lambda_-$ branch. The strong energy condition, $\sigma_0+2p_0/c^2\ge0$ ($w_0\ge-1/2$), fails only on the $\Lambda_+>\Lambda_-$ tension branch, for $-2/3< w_0<-1/2$.} ($w_0<0$, $\sigma_0>0$, $\lambda>0$) require $m_+/m_->1$ \emph{and} $\Lambda_+/\Lambda_-\ge1$, and are excluded for $\Lambda_+<\Lambda_-$.

\emph{How large is the shadow deviation?} For a supermassive black hole ($\sim10^{10}\,M_\odot$) with a Planck-scale $\Lambda$~\cite{Planck2018} and ratios $0.1\%$ from unity ($m_+/m_-=1.001$, $\Lambda_+/\Lambda_-=1.001$ or $0.999$), the transparent shell acts as a gravitational refractive layer for a distant static observer. The deviation is set mainly by the shell radius $R_0$, so at a given scale either pressure sign gives the same order: the tension shells of interest sit near the static radius and are faint ($\sim10^{-12}$), whereas positive-pressure photon-sphere shells reach $\sim10^{-3}$ even at this $0.1\%$ excess, about two orders of magnitude below current EHT precision on M87* and Sgr~A*~\cite{Galison:2023qlm,EventHorizonTelescope:2024xos,2021ChJSS..41..211K,EventHorizonTelescope:2025vum} but within sub-percent reach of next-generation arrays~\cite{EventHorizonTelescope:2024xos,2021ChJSS..41..211K,EventHorizonTelescope:2025vum,2025SCPMA..6919511H,Mroczkowski:2024htv,Akiyama:2026dhk}. The deviation depends on the exterior parameters $(m_+,\Lambda_+,r_O)$, which an outside observer can constrain, and on the hidden interior ones $(m_-,\Lambda_-,R_0)$, which it cannot. Since the shadow size is set by the interior mass $m_-$, a deviation is degenerate with simply assigning the central object a different mass; and because $(w_0,\lambda)$ enter only through $R_0$, a single measurement cannot recover the equation of state. Constraining the shell would thus need external priors on $(m_\pm,\Lambda)$~\cite{Gebhardt_2011,Walsh_2013,Planck2018} and complementary probes such as photon-ring substructure~\cite{Gralla:2019xty} or strong-field lensing~\cite{Perlick:2021aok,Perivolaropoulos:2023zzc}; the shadow alone is only a consistency test.

Our analysis covers only the radial sector. In the eikonal limit, Yang, Bonga, and
Pan~\cite{Yang:2022gic} showed that a membrane between two vacuum regions---including ones with
different cosmological constants, as here---is warping-unstable for $p/(\sigma c^2)>0$ (extended to
all angular scales in~\cite{Pitre:2026msx}) and warping-stable for $p/(\sigma c^2)<0$. Our
positive-pressure photon-sphere shells are thus warping-unstable despite their larger deviation
($\sim10^{-3}$), while the tension shells ($\sigma_0>0$, $p_0<0$) are warping-stable but faint
($\sim10^{-12}$). Radial stability is necessary but not sufficient; a full non-radial analysis,
along with rotating and charged generalizations, is left to future work. 

\begin{acknowledgments}
This research was supported by COST Action CA21136 - Addressing observational tensions in cosmology with systematics and fundamental physics (CosmoVerse), supported by COST (European Cooperation in Science and Technology). Furthermore, Dimitrios Efstratiou was supported by a scholarship from the University of Ioannina under the project ``Enhancement and Support of the Operational, Research, and Educational Activities of the University of Ioannina'' (Project Code: 82985/144059/$\beta$6.$\varepsilon$).
\end{acknowledgments}
 
\section*{Data Availability Statement}
The figure-generating Mathematica (v13) and Python (3.12.12) code will be publicly accessible in the GitHub repository and licensed under MIT in~\citep{Mathematicafile}.

\appendix

\section{Induced metric and extrinsic curvature}
\label{app:junction}
 
We adopt intrinsic coordinates $y^{i}=(c\tau,\theta,\phi)$, identical on both sides of $\Sigma$, with $\tau$ the proper time of a comoving observer. The isometric embedding $i_\pm:\Sigma\hookrightarrow\mathcal{M}^\pm$ is
\begin{equation}\label{embedding}
i^\mu_\pm(y^i)=x^\mu_\pm(y^i)=\bigl(c\,t_\pm(\tau),\,R_\pm(\tau),\,\theta,\,\phi\bigr),
\end{equation}
with $R_\pm(\tau)$ the comoving radius in each chart. The induced metric (first fundamental form) is the pull-back
\begin{equation}\label{firstff}
h^\pm_{ij}\equiv\bigl[(i^\pm)^*g^\pm\bigr]_{ij}
=\frac{\partial x^\alpha_\pm}{\partial y^i}\,\frac{\partial x^\beta_\pm}{\partial y^j}\,g^\pm_{\alpha\beta}.
\end{equation}
Continuity across $\Sigma$~\cite{Israel:1966rt,Poisson:2009pwt,GronHervik2007GR} requires
\begin{equation}\label{continuity}
h^-_{ij}=h^+_{ij}\ \Longrightarrow\ R_+=R_-\equiv R ,
\end{equation}
though the radial coordinate basis vectors $\partial_{r_\pm}$ need not coincide on $\Sigma$ (see also~\cite{Cao:2025qzy}). The induced line element is
\begin{equation}\label{eq:induced_line_element}
ds^2_\Sigma=h_{ij}\,dy^i dy^j= c^2 d\tau^2 - R(\tau)^2\bigl(d\theta^2+\sin^2\theta\,d\phi^2\bigr),
\end{equation}
so a comoving observer ($v^\theta=v^\phi=0$, $h_{ij}v^iv^j=c^2$) has $v^\tau=c$. The bulk four-velocity of a shell element is the push-forward of $v^i$, $u^\mu_\pm=[i^\pm_*(\mathbf{v})]^\mu =\frac{\partial x^\mu_\pm}{\partial y^i}\,v^i = c\,\dot t_\pm\,\delta^\mu_{\,t}+\dot R_\pm\,\delta^\mu_{\,r}$. Imposing the normalisation $u^\mu_\pm u^\pm_\mu=c^2$ determines 
\begin{equation}\label{fourvel}
\mathbf{u}_\pm=\left(\frac{c}{f_\pm}\sqrt{f_\pm+(\dot R/c)^2},\ \dot R_\pm,\ 0,\ 0\right).
\end{equation}
The outward unit normal ($u^\mu_\pm n^\pm_\mu=0$, $n^\mu_\pm n^\pm_\mu=-1$) is
\begin{equation}\label{normal}
\mathbf{n}_\pm=\left(\frac{\dot R}{c\,f_\pm},\ \sqrt{f_\pm+(\dot R/c)^2},\ 0,\ 0\right).
\end{equation}
The extrinsic curvature, as a 3-tensor on $\Sigma$,\footnote{The projector onto $\Sigma$ is $P_{\mu\nu}=g_{\mu\nu}-\epsilon\,n_\mu n_\nu$ with $\epsilon\equiv n^\mu n_\mu=-1$ (see \cite{Carroll:2004st}). The extrinsic curvature is $K_{\mu\nu}=\tfrac12\mathcal{L}_n P_{\mu\nu}
=\nabla_\mu n_\nu-\epsilon\,n_\mu a_\nu$, with $a^\mu=n^\nu\nabla_\nu n^\mu$. Its projection onto $\Sigma$ is $\hat K_{\alpha\beta}=P^{\mu}{}_{\alpha}P^{\nu}{}_{\beta}\nabla_\mu n_\nu$, and the induced 3-tensor is the pull-back $\mathcal{K}_{ij}=[\,i^*(\hat K)\,]_{ij} =\dfrac{\partial x^\mu}{\partial y^i}\dfrac{\partial x^\nu}{\partial y^j}\nabla_\mu n_\nu$.} is

\begin{equation}\label{Kdef}
\mathcal{K}^{\pm}_{ij}=\frac{\partial x^\mu_\pm}{\partial y^i}\frac{\partial x^\nu_\pm}{\partial y^j}\nabla_\mu n^\pm_\nu .
\end{equation}
Inserting \eqref{embedding}, \eqref{eq:induced_line_element}, and \eqref{normal} into the Israel conditions \eqref{israeljunc0} yields the perfect-fluid stress--energy \eqref{perfectfluid}, the junction condition \eqref{junc1main} (from the $\theta\theta$ component), and the conservation law \eqref{conservationofenergysigma1} (from $j=\tau$). Considering the $\theta\theta$  components of the extrinsic curvature, we have
\[
\mathcal{K}_{\theta\theta}^\pm = -\Gamma^\mu_{\theta\theta}\, n_\mu^\pm=f^{-1}_{\pm}\Gamma^r_{\theta\theta}n_\pm^r=-\,R \sqrt{\,f_\pm + (\dot R/c)^2\,}.
\]

 \section{Analytic approximation of the stability bounds for $\lambda=1$ in the test-shell limit}
\label{app:analytic}
 
The stability windows can be obtained in closed form in the \emph{test--shell} limit $m_+/m_-\to1$, $\Lambda_+/\Lambda_-\to1$. Introducing $\delta m\equiv m_+-m_-$, $\delta\Lambda\equiv\Lambda_+-\Lambda_-$ and
\begin{equation}\label{eq:Delta_def}
\Delta\equiv f_--f_+=\frac{2G\,\delta m}{c^2R}+\frac{\delta\Lambda}{3}R^2 ,
\end{equation}
the potential~\eqref{veffenergyeqmain} takes the exact form, with $\bar f\equiv\tfrac12(f_++f_-)$,
\begin{equation}\label{eq:Veff_exact}
V_{\rm eff}=\frac{c^2}{2}\Big[\bar f-\frac{\Delta^2}{4\kappa^2}-\frac{\kappa^2}{4}\Big].
\end{equation}
Let $e\equiv\max(\delta m/m_-,|\delta\Lambda|/\Lambda_-)\ll1$ and assume that $9G^2m_-^2\Lambda_-/c^4\ll1$. The mismatch is first order, $\Delta=O(e)$, being linear in $\delta m,\delta\Lambda$. The junction condition fixes $\kappa=\beta_--\beta_+=\Delta/(\beta_-+\beta_+)=O(e)$, since $f_\pm$ are $O(1)$ away from the horizons. Their ratio stays finite, $g\equiv\Delta/2\kappa=\tfrac12(\beta_-+\beta_+)=O(1)$. Thus $\bar f$ and $\Delta^2/4\kappa^2=g^2$ are both $O(1)$ and balance, while the self-gravity term $\kappa^2/4=O(e^2)$ is subleading. Discarding it ($f\equiv\bar f,\, g\equiv\frac{\Delta}{2\kappa}$),
\begin{equation}\label{eq:Veff_reduced}
V_{\rm eff}(R)=\frac{c^2}{2}\big[f(R)-g^2(R)\big]+O(e^2).
\end{equation}
 Because $V_{\rm eff}''(R_0)=0$ is controlled by this $O(1)$ balance, and the window edges turn only on which term dominates $\Delta$ (the ratio ordering), the closed-form bounds carry $O(e^2)$
corrections and persist at finite ratios.
 
With $u\equiv 2G\bar m/c^2R$ and $v\equiv\bar \Lambda R^2/3$, where $\bar m=(m_++m_-)/2$ and $\bar \Lambda=(\Lambda_++\Lambda_-)/2$, so $f=1-u-v$,
\begin{equation}\label{eq:PQrho}
P\equiv\frac{Rf'}{f}=\frac{u-2v}{1-u-v},\quad
Q\equiv\frac{R^2f''}{f}=\frac{-2(u+v)}{1-u-v}.
\end{equation}
Combining \eqref{conservationofenergysigma1} with \eqref{linearEoS} gives the identities
\begin{equation}\label{eq:identities}
\frac{\kappa'}{\kappa}=-\frac{1+2w}{R},\quad
\frac{dw}{dR}=-\frac{2}{R}(\lambda-w)(1+w),\quad
\Delta''=\frac{2\Delta}{R^2},
\end{equation}
and differentiating \eqref{eq:Delta_def} gives $R\Delta'+\Delta=\delta\Lambda\,R^2$, so with
$\rho\equiv R\Delta'/\Delta$
\begin{equation}\label{identitybounds}
\rho+1=\delta\Lambda\,R^2/\Delta.
\end{equation}
In particular, wherever the mass term dominates $\Delta$ (near the black hole, or for $\Lambda_+=\Lambda_-$ at every radius) $\Delta\propto1/R$ and $\rho=-1$.
 Imposing $V_{\rm eff}(R_0)=V_{\rm eff}'(R_0)=0$ on \eqref{eq:Veff_reduced} yields $g_0^2=f_0$ and $f'_0/f_0=2g_0'/g_0$; using $g_0'/g_0=\Delta_0'/\Delta_0-\kappa_0'/\kappa_0=(\rho_0+1+2w_0)/R_0$,
\begin{equation}\label{eq:equilibrium}
P_0=2(1+2w_0+\rho_0).
\end{equation}
A further differentiation gives $V_{\rm eff}''(R_0)\propto f_0''-2g_0'^2-2g_0 g_0''$; with $g_0''$ (involving $\Delta''=2\Delta/R^2$ and $dw/dR$ at $\lambda=1$) and using \eqref{eq:equilibrium}, $\mathcal{B}_0\equiv8w_0^2+(4\rho_0+2)w_0+2(\rho_0-1)$,
\begin{equation}\label{eq:master}
V_{\rm eff}''(R_0)=\frac{c^2}{2}\,\frac{f_0}{R_0^2}\!\left(Q_0-\tfrac12 P_0^2-2\,\mathcal{B}_0\right).
\end{equation}
Equation~\eqref{eq:master} is the test--shell limit of \eqref{Veffpp0} and shares its sign, so the boundary $V_{\rm eff}''(R_0)=0$ reduces to $2Q_0-P_0^2-4\mathcal{B}_0=0$. Throughout, $f_0$ remains $O(1)$ and never vanishes, since the shell never stabilizes at a horizon ($f=0$).
 
\textbf{\textit{(i) $R_0$ deep inside the static-radius scale.}} Here the mass term dominates $\Delta$,  so $\rho_0=-1$ and $P_0=4w_0$. With $v_0\ll u_0$, $Q_0=-2P_0$ and $\mathcal B_0=8w_0^2-2w_0-4$, $V_{\rm eff}''(R_0)=\frac{2c^2 f_0}{R_0^2}(1-2w_0)(3w_0+2)$, positive for $-\tfrac23<w_0<\tfrac12$. Through $u_0=4w_0/(1+4w_0)$, $w_0=\tfrac12$ gives $u_0=\tfrac23$, i.e.\ $R_0\simeq3Gm_-/c^2$ (the photon sphere), while $w_0=-\tfrac23$ gives $u_0=\tfrac85>1$ (inside the horizon, discarded). Attractive gravity ($f_0'>0$) makes $P_0>0$, hence $w_0>0$. A shell at the attractive-gravity scale is therefore stable for $0<w_0<\tfrac12$, with $f_0\to\tfrac13\sim O(1)$ as $w_0\to\tfrac12$.
 
\textbf{\textit{(ii) $R_0$ far outside the static-radius scale, $\Lambda_+=\Lambda_-$.}} Then $\delta\Lambda=0$, so $\rho=-1$ at all radii and $P_0=4w_0$. With $u_0\ll v_0$, $Q_0=P_0$ and $\mathcal B_0=8w_0^2-2w_0-4$, $V_{\rm eff}''(R_0)=-\frac{4c^2 f_0}{R_0^2}(3w_0^2-w_0-1)$, positive for $(1-\sqrt{13})/6<w_0<(1+\sqrt{13})/6$. The lower root sets $v_0=(5-\sqrt{13})/3$, so $f_0\simeq1-v_0\gtrsim0.54\sim O(1)$, inside the cosmological horizon. Repulsive gravity ($f_0'<0$) forces $P_0<0$, hence $w_0<0$, giving the stable window $(1-\sqrt{13})/6<w_0<0$.
 
\textbf{\textit{(iii) $R_0$ at the static-radius scale, $\Lambda_+>\Lambda_-$.}} Both $u_0,v_0\ll1$, so $P_0,Q_0$ are negligible against $\mathcal B_0=-2(3w_0+2)$, giving $V_{\rm eff}''(R_0)=\frac{2c^2 f_0}{R_0^2}(3w_0+2)$, positive for $w_0>-\tfrac23$. With $P_0$ negligible, \eqref{eq:equilibrium} fixes $\rho_0=-1-2w_0$; since $\delta\Lambda>0$, \eqref{identitybounds} gives $-2w_0=\delta\Lambda R_0^2/\Delta_0$, and $\Delta_0>0$ forces $w_0<0$. The stable window is $-\tfrac23<w_0<0$.
 
\textbf{\textit{(iv) $R_0$ at the static-radius scale, $\Lambda_+<\Lambda_-$.}} Again $u_0,v_0\ll1$, so $V_{\rm eff}''(R_0)=\frac{2c^2 f_0}{R_0^2}(3w_0+2)$ is positive for $w_0>-\tfrac23$. Now $\delta\Lambda<0$ forces $-2w_0=\delta\Lambda R_0^2/\Delta_0<0$ with $\Delta_0>0$, i.e.\ $w_0>0$: no outer negative-pressure branch exists. Stability holds for all $w_0>0$.

Only the evaluation of $\mathcal{B}_0$ used $\lambda=1$. Repeating that step
with \eqref{eq:identities} at general $\lambda$, and
inserting it into \eqref{eq:master} with the branch values of
$(P_0,Q_0,\rho_0)$ of cases (i)--(iv) yields the stability windows for any
$\lambda$ by the same steps.

\section{Full expressions for $V_{\rm eff}$ and its derivatives at $R_0$}
\label{app:explicit}
 
\begin{figure*}
    \centering
    \begin{minipage}[t]{0.48\textwidth}
        \vspace{.75cm}                       
        \centering
        \includegraphics[width=1\linewidth]{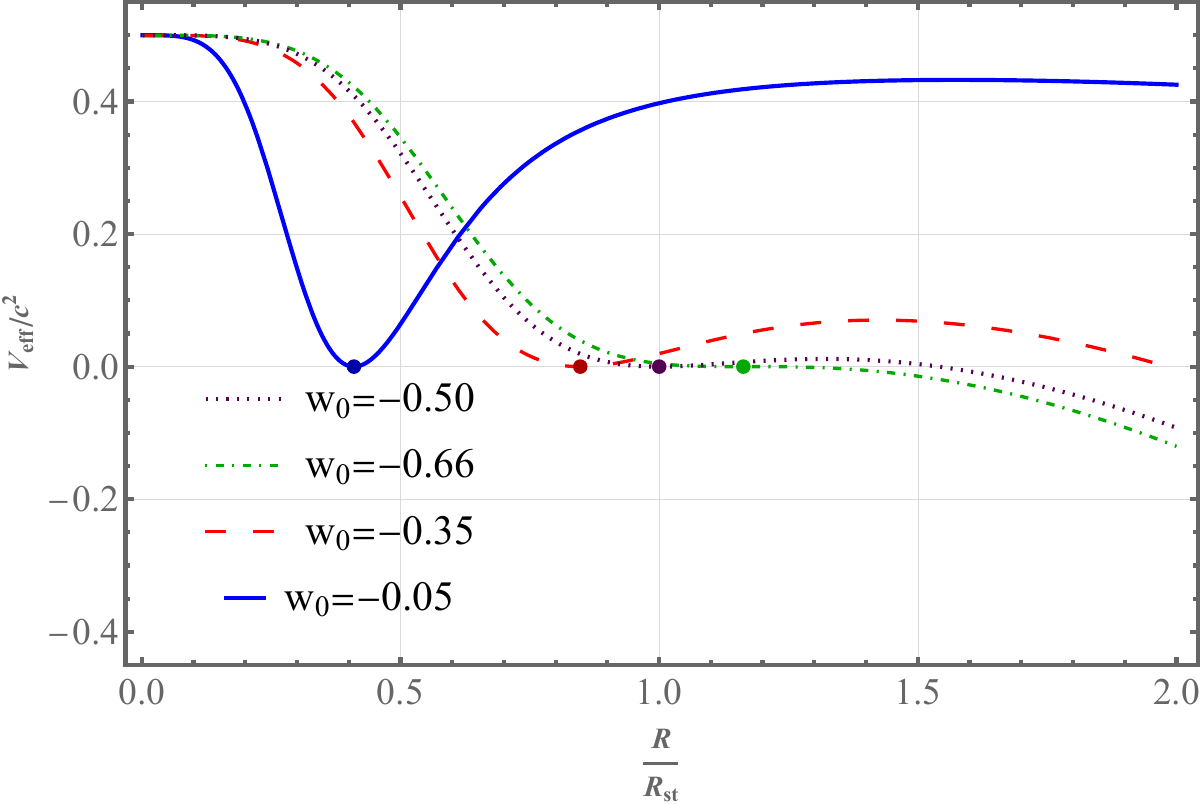}\\[-0.2ex]
        \includegraphics[width=1\linewidth]{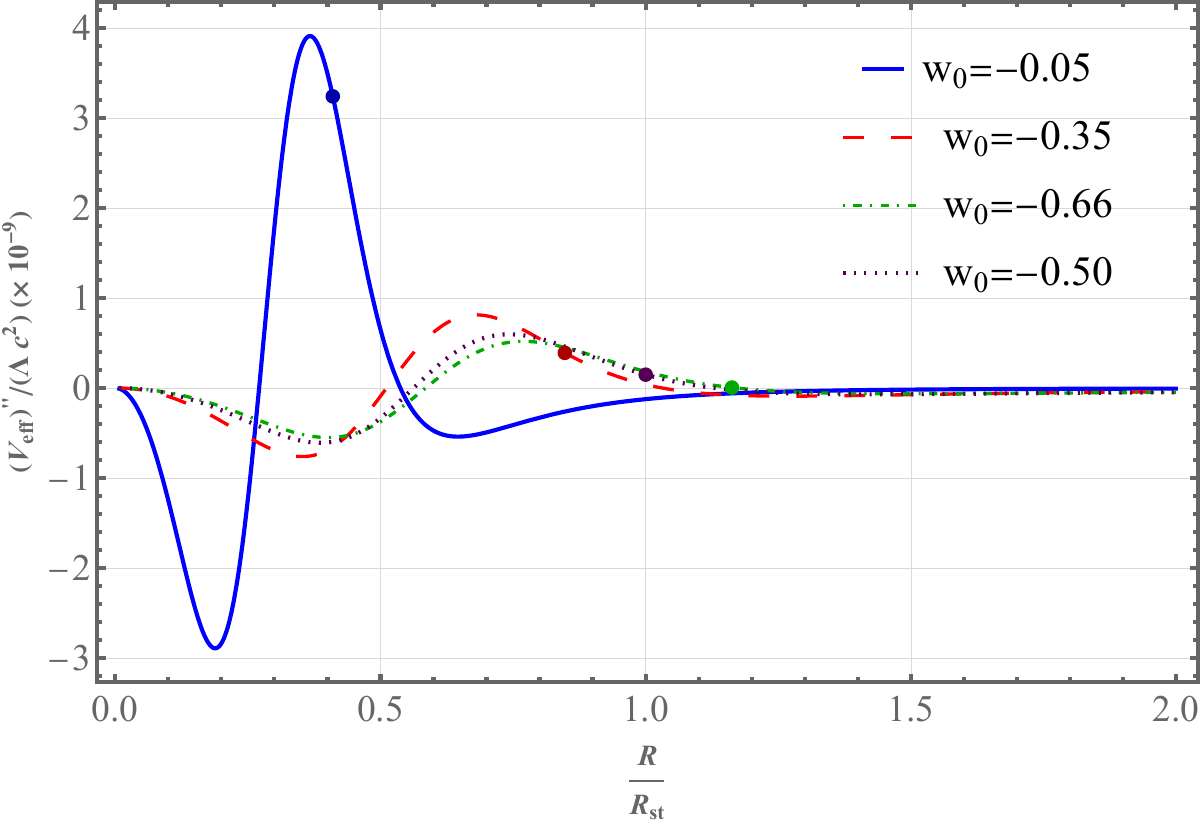}\\[-0.2ex]
        \includegraphics[width=1\linewidth]{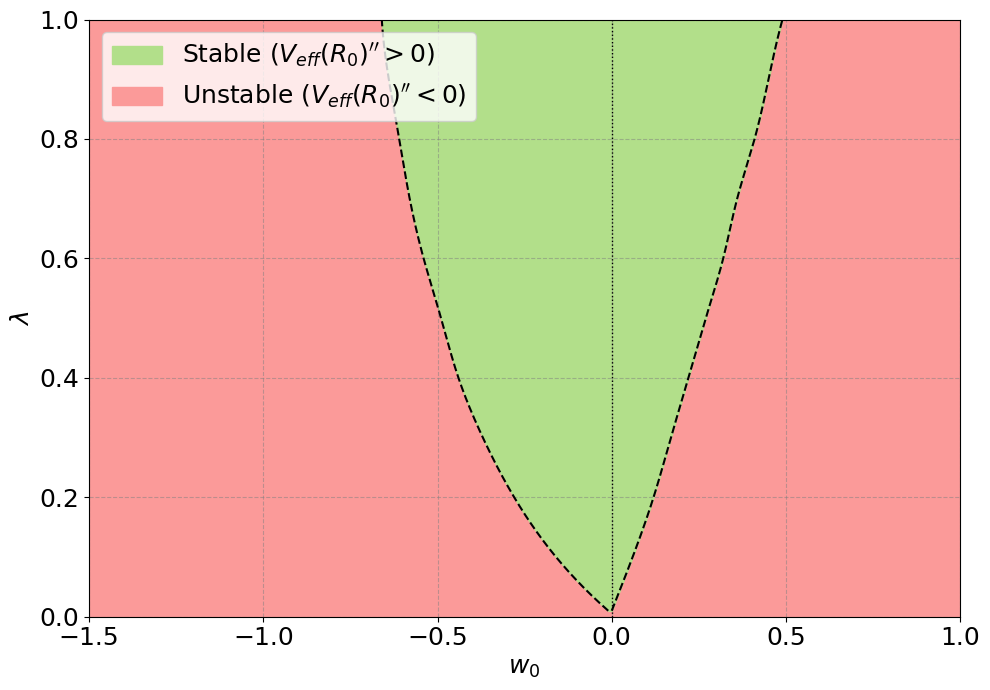}
        \captionof{figure}{Effective potential, its second derivative, and the stability map, with $m_-=10^{10}\,M_\odot$, $\rho_{\Lambda_+}=\rho_{\rm crit}=3H_0^2/(8\pi G)$~\cite{Planck2018}, $m_+/m_-=\Lambda_+/\Lambda_-=1.001$, and $\lambda=1$, giving $R_{\rm st}\equiv[3Gm_-/\bar \Lambda c^2]^{1/3}\simeq6.52\times10^{21}\,\mathrm{m}$ ($\bar \Lambda\equiv(\Lambda_++\Lambda_-)/2$). \textbf{Top:} $V_{\rm eff}/c^2$ versus $R/R_{\rm st}$ [Eq.~\eqref{veffenergyeqmain}]; dots mark $R_0$ for each $w_0$, and $V_{\rm eff}>0$ is forbidden by Eq.~\eqref{energymain} ($\dot R^2<0$). \textbf{Middle:} $V''_{\rm eff}/(\Lambda c^2)\,(\times10^{-9})$ versus $R/R_{\rm st}$; equilibria with $V''_{\rm eff}(R_0)>0$ (local minima) are stable. Since $p=\lambda(\sigma-\sigma_1)c^2$ is a local (near-equilibrium) expansion, only the neighborhood of each marked $R_0$ governs stability; elsewhere the curve extrapolates this law and is illustrative rather than reliable. \textbf{Bottom:} stable region in the $w_0$--$\lambda$ plane for the same ratios. Stable shells sit near $R_{\rm st}$ for $-2/3<w_0<0$ ($w_0=-2/3$ the boundary); for $0.01\lesssim w_0<0.5$ they cluster at $R\sim10^{13\text{--}14}\,\mathrm{m}$, approaching the photon sphere $3Gm_-/c^2\simeq4.43\times10^{13}\,\mathrm{m}$ at $w_0=0.5$.}
        \label{figvstabandsigma}
    \end{minipage}\hfill
    \begin{minipage}[t]{0.48\textwidth}
        \vspace{0pt}                       
        \centering
        \includegraphics[width=\linewidth]{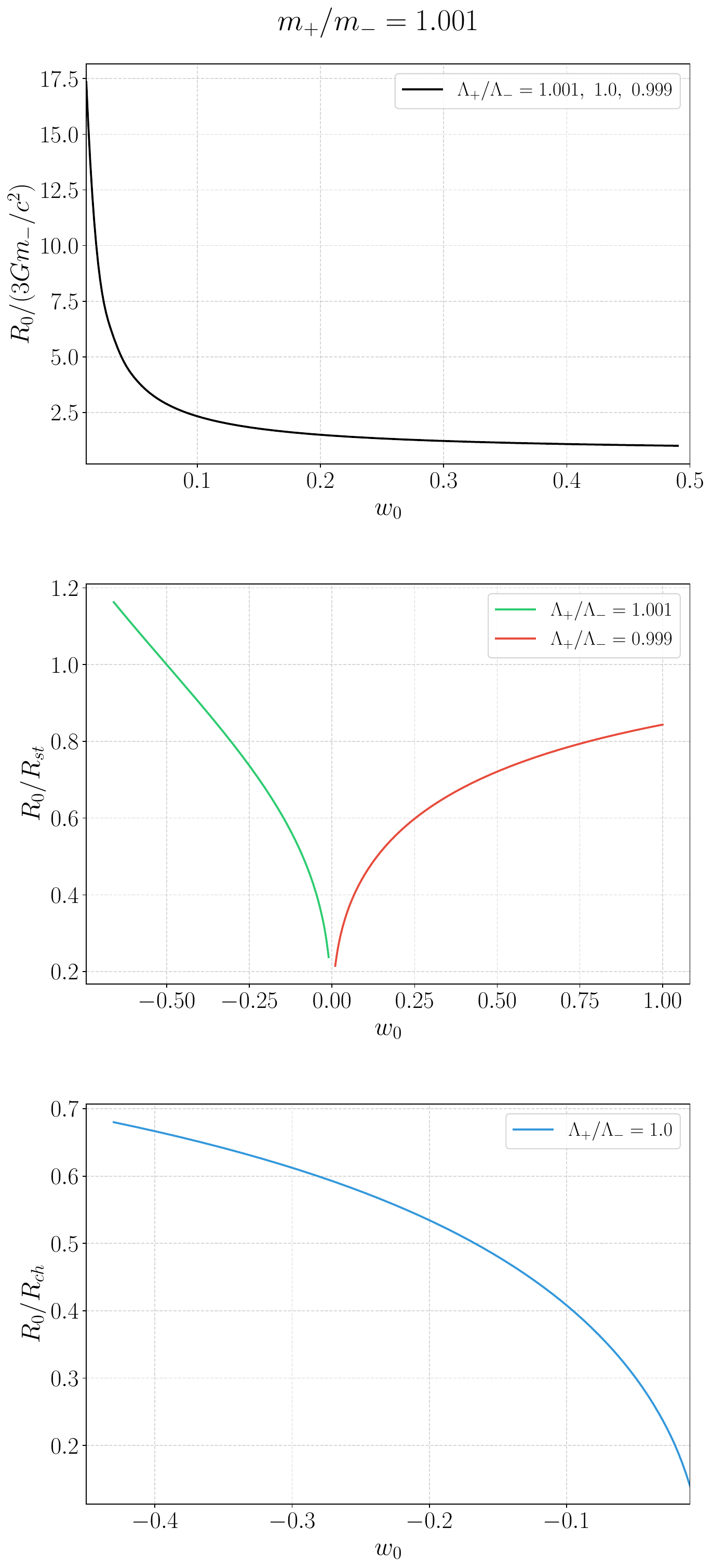}
        \captionof{figure}{Normalized stable shell radius $R_0$ versus $w_0$, for $\lambda=1$, $m_+/m_-=1.001$, $m_-=10^{10}\,M_\odot$, and $\rho_{\Lambda_+}=\rho_{\rm crit}$. Panels group the stable branches by scale. \textbf{Top:} positive-pressure shells in units of the photon-sphere radius $3Gm_-/c^2$; the ratios $\Lambda_+/\Lambda_-=1.001,\,1.0,\,0.999$ nearly coincide and approach the photon sphere as $w_0\to1/2$. \textbf{Middle:} branches at the static-radius scale $R_{\rm st}$---the negative-pressure branch for $\Lambda_+/\Lambda_-=1.001$ (bounded at $w_0=-2/3$) and the positive-pressure branch for $\Lambda_+/\Lambda_-=0.999$  ($0\lesssim w_0$). \textbf{Bottom:} the negative-pressure branch for $\Lambda_+/\Lambda_-=1.0$ in units of the cosmological horizon $R_{\rm ch}$, migrating outward as $w_0$ grows more negative.}
        \label{fig:three_cases_comparison}
    \end{minipage}
\end{figure*}
Substituting \eqref{frmetricmain}, \eqref{kapparmain}, and \eqref{EqsigmaR} into
\eqref{veffenergyeqmain}, the effective potential is
\begin{widetext}
\begin{align}
\frac{V_{\rm eff}(R)}{c^2} = \frac{1}{4} \Bigg\{ &
2 - \frac{2 G m_{-}}{c^2 R} - \frac{2 G m_{+}}{c^2 R} - \frac{R^2 \Lambda_{-}}{3} - \frac{R^2 \Lambda_{+}}{3} \notag\\
& - \frac{(1 + \lambda)^2 \left[ 6 G (m_{-} - m_{+}) + c^2 R^3 (\Lambda_{-} - \Lambda_{+}) \right]^2}
       {288 G^2 \pi^2 \sigma_0^2 \left[ R_0^2 \left( \frac{R_0}{R} \right)^{2 \lambda} (1 + w_0) + R^2 (\lambda - w_0) \right]^2}
- \frac{8 G^2 \pi^2 \sigma_0^2 R^2 \left[ -w_0 + \left( \frac{R_0}{R} \right)^{2 (1 + \lambda)} (1 + w_0) + \lambda \right]^2}
       {c^4 (1 + \lambda)^2}
\Bigg\}.
\label{veff_full}
\end{align}
\end{widetext}
Imposing $V_{\rm eff}(R_0)=V_{\rm eff}'(R_0)=0$ in the general case, with $\Lambda_+$ and $\Lambda_-$ possibly distinct, gives the equilibrium system
\begin{widetext}
\begin{equation}\label{systemA}
V_{\text{eff}}(R_0) =
-\frac{G\,(m_{-} + m_{+})}{2R_0}
-\frac{c^2}{12}\left(-6 + R_0^2(\Lambda_{-} + \Lambda_{+})\right)
-\frac{\left(6cG(m_{-} - m_{+}) + c^3 R_0^3(\Lambda_{-} - \Lambda_{+})\right)^2}
{1152\,G^2\pi^2 R_0^4 \sigma_0^2}
-\frac{2G^2\pi^2 R_0^2 \sigma_0^2}{c^2}=0,
\end{equation}
\begin{equation}\label{systemB}
V'_{\text{eff}}(R_0) =
-\frac{1}{576\,c^2 G^2 \pi^2 R_0^5 \sigma_0^2}\Bigg[
6 c^6 G (m_{-} - m_{+}) R_0^3 (3 + 4 w_0)(\Lambda_{-} - \Lambda_{+})
+ c^8 R_0^6 (3 + 2 w_0)(\Lambda_{-} - \Lambda_{+})^2-
\end{equation}
\begin{equation*}
\quad - 288 c^2 G^3 (m_{-} + m_{+}) \pi^2 R_0^3 \sigma_0^2
- 2304 G^4 \pi^4 R_0^6 (1 + 2 w_0)\sigma_0^4
+ 24 c^4 G^2 \left(3 (m_{-} - m_{+})^2 w_0 + 4 \pi^2 R_0^6 (\Lambda_{-} + \Lambda_{+}) \sigma_0^2\right)
\Bigg]=0 .
\end{equation*}
\end{widetext}
Stability requires $V_{\mathrm{eff}}''|_{R_0}>0$, with (see Fig.~\ref{figvstabandsigma})
\begin{widetext}
\begin{equation}\label{Veffpp0}
\begin{split}
V_{\text{eff}}''(R_0) =
& - \frac{G (m_{-} + m_{+})}{R_0^3}
+ \frac{c^2}{24} \Bigg[
    - 4 (\Lambda_{-} + \Lambda_{+})
    - \frac{3 (m_{-} - m_{+})^2 \left( w_0 (1 + 6 w_0) - 2 (1 + w_0) \lambda \right)}{\pi^2 R_0^6 \sigma_0^2}
\Bigg] \\
& - \frac{c^6 (15 + 26 w_0 + 12 w_0^2 - 4 (1 + w_0) \lambda) (\Lambda_{-} - \Lambda_{+})^2}{576 G^2 \pi^2 \sigma_0^2}
+ \frac{c^4 (m_{-} - m_{+}) \left[ 3 + 2 w_0 (7 + 6 w_0 - 2 \lambda) - 4 \lambda \right] (-\Lambda_{-} + \Lambda_{+})}{48 G \pi^2 R_0^3 \sigma_0^2} \\
& - \frac{4 G^2 \pi^2 (3 + 6 w_0 + 4 w_0^2 + 4 (1 + w_0) \lambda) \sigma_0^2}{c^2}.
\end{split}
\end{equation}
\end{widetext}
The equilibria follow from the system [Eqs.~\eqref{systemA}--\eqref{systemB}], and their stability from the sign of $V_{\rm eff}''(R_0)$ [Eq.~\eqref{Veffpp0}].

Figure~\ref{fig:three_cases_comparison} shows how the stable equilibrium radius $R_0$ varies with $w_0$, at $\lambda=1$ and $m_+/m_-=1.001$. Each panel collects the branches that sit on a common scale and normalises $R_0$ to it, so that the curves within a panel are directly comparable. Top: the photon-sphere branch, in units of $3Gm_-/c^2$. It exists for all three orderings $\Lambda_+/\Lambda_-=1.001,\,1.0,\,0.999$, the curves nearly coinciding, and $R_0$ falls steadily with $w_0$, reaching the photon sphere only as $w_0\to\tfrac12$. Middle: the two branches at the static-radius scale, in units of $R_{\rm st}$. For $\Lambda_+>\Lambda_-$ the tension branch moves outward as $w_0$ decreases, ending at its bound $w_0=-2/3$; for $\Lambda_+<\Lambda_-$ a positive-pressure branch ($0\lesssim w_0$) moves outward as $w_0$ increases, both reaching their smallest $R_0$ near $w_0=0$. Over $0\lesssim w_0 \lesssim  \tfrac12$ this second branch coexists with a photon-sphere equilibrium, plotted in the top panel: at the same $w_0$ the shell may sit stably at either scale. Bottom: the $\Lambda_+=\Lambda_-$ tension branch, in units of the cosmological horizon $R_{\rm ch}$, moving outward as $w_0$ grows more negative. Static stable shells are thus possible at several scales, with $R_{\rm st}\equiv\left[3Gm_-/\bar\Lambda c^2\right]^{1/3}$ ($\bar \Lambda\equiv(\Lambda_++\Lambda_-)/2$) separating an inner, positive-pressure range from an outer, negative-pressure one whenever $\Lambda_+\geq\Lambda_-$; for $\Lambda_+<\Lambda_-$ the tension branch is absent and positive-pressure shells occupy both scales. Here $R_{\rm st}$ denotes only a characteristic scale and does not retain its usual SdS interpretation (see Appendix~\ref{sec:appendixa}).

\section{The Schwarzschild--de Sitter shadow for a static observer}
\label{sec:appendixa}

For a static, spherically symmetric black hole of mass $m$ with positive $\Lambda$, the vacuum solution $R_{\mu\nu}-\Lambda g_{\mu\nu}=0$ is the SdS metric with $f(r)=1-2Gm/(c^2r)-\Lambda r^2/3$. Demanding that $f(r)=0$ have two positive roots and one negative root gives two physical horizons: the smaller positive root is the black-hole horizon---close to (for $\rho_\Lambda=\rho_{\rm crit}$~\cite{Planck2018}) but slightly outside the Schwarzschild radius---and the larger defines the cosmological horizon~\cite{GibbonsHawking1977,Bousso2002}. A key feature is the radius where attraction and cosmological repulsion balance~\cite{Stuchlik1999,Balaguera2005}: for radial timelike geodesics ($h=0$), $\ddot r=-c^2(Gm/c^2r^2-\Lambda r/3)$~\cite{Hobson:2006} vanishes at $R_{\rm st}\equiv(3Gm/\Lambda c^2)^{1/3}$, where a particle released from rest stays static.

\subsection{Lightlike trajectories in Schwarzschild-de Sitter spacetime}

We parameterize null geodesics by an affine $\zeta$ ($\dot{}\equiv d/d\zeta$) in the equatorial plane $\theta=\pi/2$. The null condition $ds^2=0$ gives the conserved quantities $\dot t=k/f(r)$ and $\dot\phi=h/r^2$ ($k,h$ constants) and $c^2k^2=\dot r^2+(h^2/r^2)f(r)$~\cite{Hobson:2006}; with the impact parameter $b\equiv h/(ck)$ the orbit-shape equation follows, 

\begin{equation}\label{photonshape} 
\left(\frac{dr}{d\phi}\right)^2=\frac{1}{b^2}r^4-r^2f(r),
\end{equation}
so the spatial trajectory $r(\phi)$ is fixed by $b$ and $f(r)$ alone.

Now, by introducing the substitution \( u \equiv 1/r \) into Eq.~\eqref{photonshape} and then differentiating with respect to the azimuthal angle \(\phi\), we obtain the second-order differential equation $\frac{d^2u}{d\phi^2} = \frac{3Gm}{c^2} \, u^2 - u $. Setting $u=u_0$ and $du/d\phi=d^2u/d\phi^2=0$ we obtain the radius of the circular photon orbit as $ r_0=\frac{3Gm}{c^2}$. The circular light ray at \( r = 3Gm/c^{2} \) is unstable. A small perturbation of the photon's initial trajectory in the equatorial plane will cause it to depart from this circular path and ultimately cross either the inner or outer horizon. Moreover, we rewrite Eq. (\ref{photonshape}) as
\begin{equation}\label{orbiteq}
    \frac{dr}{d\phi} =\pm  
     r^{2} \left( \frac{1}{b^{2}} - \frac{1}{r^{2}}+\frac{2Gm}{c^2r^3}+\frac{\Lambda}{3}\right)^{1/2}.
\end{equation}
The critical value $b = b_{\rm crit}$ corresponds to the threshold between trajectories that plunge into the black hole and those that scatter away.  Note that if we set as $r=r_0$ then Eq. (\ref{orbiteq}) yields
\begin{equation}\label{criticalimpactparameterapp}
    b_{\rm crit}=\frac{3\sqrt{3}Gm}{c^2\sqrt{1-\frac{9G^2m^2\Lambda}{c^4}}}.
\end{equation}
 
\subsection{Shadow of the SdS black hole for a static observer}

 For a static observer $O$, at $\theta_O=\pi/2$, the four-velocity has only a  $t-$component, $ [u_O^\mu] = (u_O^t,0,0,0)$\cite{PhysRevD.60.044006}. Imposing the normalization condition $g_{\mu\nu}u^\mu u^\nu = c^2$ yields $ g_{tt}(u_O^t)^2 = c^2 \quad \Rightarrow \quad f(r)(u_O^t)^2 = c^2$, from which we obtain $ u_O^t = \frac{c}{\sqrt{f(r_O)}}$.
    The observer’s orthonormal tetrad vectors \cite{Carroll:2004st,Hobson:2006} are chosen as
$(\hat e_{\hat t})^\mu = u_O^\mu/c$, 
    $(\hat e_{\hat r})^\mu = \left(0,\sqrt{f(r_O)},0,0\right),$
together with the angular basis vectors
 $ (\hat e_{\hat \theta})^\mu = \left(0,0,r_O^{-1},0\right)$, $(\hat e_{\hat \phi})^\mu = \left(0,0,0,(r_O\sin\theta_O)^{-1}\right)$.
Note that the metric satisfies 
$g_{\mu\nu}(\hat{e}_{a})^{\mu}(\hat{e}_{b})^{\nu}=\eta_{\hat a \hat b}$
with \([\eta_{ \hat a \hat b}]=\operatorname{diag}(1,-1,-1,-1)\).

When the observer's worldline intersects that of a photon confined to the equatorial plane $\theta=\pi/2$ with coordinate four-momentum $[p^\mu]=\left(k/f,\;\dot r,\;0,\;h/r^2\right)$, at an event $P$, projecting the photon four-momentum onto the static observer's orthonormal tetrad at $r=r_O$ gives the following.
The projections of $p^\mu$ are then $p_{\hat t} = g_{\mu\nu}\,p^\mu\,(\hat e_{\hat t})^\nu = k/\sqrt{f(r_O)}$, $p_{\hat r} = g_{\mu\nu}\,p^\mu\,(\hat e_{\hat r})^\nu$ and $p_{\hat\phi} = g_{\mu\nu}\,p^\mu\,(\hat e_{\hat\phi})^\nu=- h/r_O$.
Note that the observer measures the photon direction from the spatial 3-vector \(\mathbf{p}=(p^{\hat r},0,p^{\hat\phi})\). The angular radius \(\theta_*\) of the shadow (angle between the photon's spatial direction and the radial inward direction) may be taken from the ratio of the orthonormal spatial components  $\tan\theta_*=\frac{|p^{\hat\phi}|}{|p^{\hat r}|}$.
Using the projected components above  we get
$\tan\theta_*
=  \frac{h\sqrt{f(r_O)}}{r_O\,|\dot r|_{r=r_O}}$.
Writing \(dr/d\varphi = \dot r/\dot\varphi = \dot r/(h/r^2)\) we have
$|\dot r| = \frac{h}{r^2}\left|\frac{dr}{d\varphi}\right|$,
and substituting into the above yields 
$\tan^2\theta_*
=r_O^2 f(r_O)/\left(dr/d\varphi\right)^2_{r=r_O}$.
We substitute the Eq. (\ref{orbiteq}) and we get
$\tan^2\theta_*
= \frac{f(r_O)}
{\frac{1}{b^2}r_O^2-f(r_O)}$.
Using
\(\sin^2\theta=\dfrac{\tan^2\theta}{1+\tan^2\theta}\) one finds after simplifying 
$\sin^2\theta_*=\frac{f(r_O)\,b^{\,2}}{r_O^2}$.
Substituting the corresponding critical impact parameter (Eq. (\ref{criticalimpactparameterapp})), we obtain the angular radius of the shadow as measured by a static observer at \(r_{O}\) \cite{PhysRevD.60.044006,Perlick:2018iye,Perlick:2021aok}:
\begin{equation}\label{eq:theta-final}
\sin^2\theta_{\rm crit}
\;=\frac{1-\frac{2Gm}{c^2r_O}-\frac{\Lambda r_O^2}{3}}{r_O^2\left(\frac{c^4}{27G^2m^2}-\frac{\Lambda}{3}\right)}.
\end{equation}
For the static observer, the black hole shadow occupies an angular radius $\theta_{\rm crit}$ in the sky. This radius decreases with increasing coordinate distance $r_O$: at the horizon itself, it encompasses the full sky; at the photon sphere, it subtends exactly half the sky ($\theta_{\rm crit}=\pi/2$); and it diminishes  to zero as the static observer approaches the cosmic horizon \cite{PhysRevD.60.044006}.

\bibliographystyle{apsrev4-2}
\bibliography{references}

\end{document}